\documentclass[sigconf,]{acmart}
\PassOptionsToPackage{bookmarks=false}{hyperref}
\pdfoutput=1

\AtBeginDocument{%
  \providecommand\BibTeX{{%
    \normalfont B\kern-0.5em{\scshape i\kern-0.25em b}\kern-0.8em\TeX}}}

\usepackage[ruled,vlined,lined,commentsnumbered]{algorithm2e}
\usepackage{booktabs}
\usepackage{moreverb}
\usepackage{fontenc}
\usepackage{amsmath}

\usepackage{amssymb}
\usepackage{fancybox}
\usepackage{color}
\usepackage{colortbl}
\usepackage{array}
\usepackage{diagbox}
\usepackage{flushend}
\usepackage{multirow}
\usepackage{multicol}
\usepackage{listings}
\usepackage{makecell}
\usepackage{graphicx}
\usepackage{setspace}
\usepackage{soul}
\usepackage[T1]{fontenc}
\usepackage{csvsimple}
\usepackage{longtable}
\usepackage[skip=0pt,font=small,labelfont=bf]{caption}
\usepackage{url}
\usepackage{lscape}
\usepackage{rotating}
\usepackage{tikz}
\usepackage{enumitem}
\usepackage{subcaption}
\usepackage{wrapfig}
\usepackage[most]{tcolorbox}
\usepackage{microtype}
\makeatletter
\@namedef{ver@lineno.sty}{9999/12/31}
\@namedef{opt@lineno.sty}{}
\makeatother
\usepackage[frozencache,cachedir=minted-cache]{minted}
\definecolor{wtbg}{rgb}{1.0,1.0,1.0}
\definecolor{rdbg}{rgb}{0.965,0.75,0.74}
\definecolor{grbg}{rgb}{0.73,0.85,0.65}
\newmintinline{perl}{python3, fontsize=\normalsize, bgcolor=grbg,numbersep=0pt,xleftmargin=0em,frame=none,}
\newmintinline{python}{python3, fontsize=\normalsize, numbersep=0pt,xleftmargin=0em,frame=none}
\usepackage{threeparttable}

\usepackage{marvosym}
\usepackage{pifont}% http://ctan.org/pkg/pifont%\usepackage{subfigure}

\usepackage{arydshln}
\usepackage{booktabs}

\newcolumntype{L}[1]{>{\raggedright\let\newline\\\arraybackslash\hspace{0pt}}m{#1}}
\newcolumntype{C}[1]{>{\centering\let\newline\\\arraybackslash\hspace{0pt}}m{#1}}
\newcolumntype{R}[1]{>{\raggedleft\let\newline\\\arraybackslash\hspace{0pt}}m{#1}}

\definecolor{codegreen}{rgb}{0,0.6,0}
\definecolor{codegray}{rgb}{0.5,0.5,0.5}
\definecolor{codepurple}{rgb}{0.58,0,0.82}
\definecolor{backcolour}{rgb}{0.95,0.95,0.92}
\definecolor{lightgreen}{HTML}{99d8c9}

\lstdefinestyle{mystyle}{
    commentstyle=\color{codegreen},
    keywordstyle=\color{magenta},
    numberstyle=\tiny\color{black},
    stringstyle=\color{codepurple},
    basicstyle=\footnotesize,
    breakatwhitespace=false,
    breaklines=true,
    captionpos=b,
    keepspaces=true,
    showspaces=false,
    showstringspaces=false,
    showtabs=false,
    tabsize=2
}

\setlist{noitemsep} %to leave space around whole list

\definecolor{darkpastelred}{rgb}{0.76, 0.23, 0.13}
\definecolor{ao(english)}{rgb}{0.0, 0.5, 0.0}
\definecolor{diffgray}{RGB}{187,187,187}
\definecolor{diffred}{RGB}{161,0,0}
\definecolor{diffgreen}{RGB}{0,133,0}

\definecolor{darkpastelred}{rgb}{0.76, 0.23, 0.13}
\definecolor{ao(english)}{rgb}{0.0, 0.5, 0.0}
\lstdefinelanguage{diff}{
	morecomment=[f][\color{blue}]{@@},     % group identifier
	morecomment=[f][\color{red}]-,         % deleted lines
	morecomment=[f][\color{codegreen}]+,       % added lines
	morecomment=[f][\color{red}]{---}, % Diff header lines (must appear after +,-)
	morecomment=[f][\color{codegreen}]{+++},
}

\definecolor{yellow}{RGB}{255,255,153}
\definecolor{grey}{RGB}{224,224,224}

\newboolean{showcomments}
\setboolean{showcomments}{true}
\ifthenelse{\boolean{showcomments}}
 { \newcommand{\mynote}[2]{
      \fbox{\bfseries\sffamily\scriptsize#1}
        {\small$\blacktriangleright$\textsf{\emph{#2}}$\blacktriangleleft$}}}
        { \newcommand{\mynote}[2]{}}

\definecolor{DarkOrange}{rgb}{0.8,0.3,0.0}
\definecolor{DarkCyan}{rgb}{0.0, 0.55, 0.55}

\newcolumntype{?}{!{\vrule width 1pt}}

\definecolor{grey}{rgb}{0.9,0.9,0.9}
\definecolor{lightgrey}{HTML}{f0f0f0}
\definecolor{mygreen}{HTML}{02818a}
\definecolor{mygray}{HTML}{666666}

\newcommand{\toolname}{\texttt{Code\-Guard\-er}\xspace}
\newcommand*{\eg}{e.g., }

\newcommand*{\ie}{i.e., }

\newcommand{\etal}{\emph{et~al.}\xspace}

\newcommand{\update}[1]{\textcolor{blue}{#1}}

\newcounter{myboxcounter}
\newcommand{\prompt}[3]{
\refstepcounter{myboxcounter}
\begin{tcolorbox}[colback=gray!10, colframe=black!80,
width=\linewidth, arc=2mm, auto outer arc, title={{\small #1}}, label={prompt:#2}, center, left=2mm,right=2mm]
{
\begingroup
\small
\vspace{-2mm}
\linespread{0.7}\selectfont
#3
\vspace{-2mm}
\endgroup
}
\end{tcolorbox}
}

\newcommand{\notez}[1]{
\begin{tcolorbox}[size=fbox,boxrule=0.5pt,top=0.5pt,bottom=0.5pt,
colframe=blue!5!black,colback=black!5!white]
\em #1
\end{tcolorbox}
}

\begin{document}

% paper title
\title{Give LLMs a Security Course: Securing Retrieval-Augmented Code Generation via Knowledge Injection}
%\subtitle{A Systematic Assessment of 16 Automated Repair Systems for Java Programs}
% \thanks{$^\dagger$ Shangwen Wang is the corresponding author. \\
% Bo Lin, Shangwen Wang, and Xiaoguang Mao are with the Key Laboratory of Software Engineering for Complex Systems.\\
% Yepang Liu is with the Research Institute of Trustworthy Autonoumous Systems.
% }

% \settopmatter{printacmref=false}

\author{Bo Lin}
%\authornotemark[2]
\email{linbo19@nudt.edu.cn}
\authornote{Bo Lin, Shangwen Wang, Yihao Qin, Liqian Chen and Xiaoguang Mao are also with the State Key Laboratory of Complex \& Critical Software Environment}

\affiliation{%
  \institution{College of Computer Science, National University of Defense Technology}
% %  \streetaddress{P.O. Box 1212}
 	\country{Changsha, China}
% %  \state{Ohio}
% %  \postcode{43017-6221}
}

\author{Shangwen Wang}
% \authornote{The first two authors contributed equally to this work, and Ming Wen is the corresponding author.}
%\authornote{College of Computer Science, NUDT, Changsha, 410073, China}
% \authornotemark[1]
\email{wangshangwen13@nudt.edu.cn}
\authornote{Shangwen Wang is the corresponding authors.}
\affiliation{%
  \institution{College of Computer Science, National University of Defense Technology}
 %  \streetaddress{P.O. Box 1212}
 	\country{Changsha, China}
 %  \state{Ohio}
 %  \postcode{43017-6221}
}

% \author{Ming Wen}
% \authornote{Co-first and corresponding author.}
%\authornote{Corresponding authors.}
% \authornotemark[1]
%\authornote{School of Cyber Science and Engineering, HUST, Wuhan, 430074, China}
% \authornote{National Engineering Research Center for Big Data Technology and System, Services Computing Technology and System Lab, Hubei Engineering Research Center on Big Data Security, HUST, Wuhan, 430074, China}
% \email{mwenaa@hust.edu.cn}
% \affiliation{%
%   \institution{School of Cyber Science and Engineering, Huazhong University of Science and Technology}
%  %  \streetaddress{P.O. Box 1212}
%  	\country{Wuhan, China}
%  %  \state{Ohio}
%  %  \postcode{43017-6221}
% }

% \author{Hongjun Wu}
% %\authornotemark[2]
% \email{wuhongjun15@nudt.edu.cn}
% \affiliation{
%     \institution{College of Computer Science, National University of Defense Technology}
%     \country{Changsha, China}
% }

\author{Yihao Qin}
%\authornotemark[2]
\email{qinyihao@nudt.edu.cn}
\affiliation{%
  \institution{College of Computer Science, National University of Defense Technology}
 %  \streetaddress{P.O. Box 1212}
 	\country{Changsha, China}
 %  \state{Ohio}
 %  \postcode{43017-6221}
}

% \author{Deqing Zou}
% \authornotemark[2]
% %\authornotemark[4]
% \authornote{Shenzhen HUST Research Institute, Shenzhen, 518057, China}
% \email{deqingzou@hust.edu.cn}
% \affiliation{%
%   \institution{School of Cyber Science and Engineering, Huazhong University of Science and Technology}
%  %  \streetaddress{P.O. Box 1212}
%  	\country{Wuhan, China}
%  %  \state{Ohio}
%  %  \postcode{43017-6221}
% }

\author{Liqian Chen}
% \authornotemark[1]
\email{lqchen@nudt.edu.cn}
\affiliation{%
  \institution{College of Computer Science, National University of Defense Technology}
 %  \streetaddress{P.O. Box 1212}
 	\country{Changsha, China}
 %  \state{Ohio}
 %  \postcode{43017-6221}
}

\author{Xiaoguang Mao}
% \authornotemark[1]
%\authornotemark[2]
\email{xgmao@nudt.edu.cn}
\affiliation{%
  \institution{College of Computer Science, National University of Defense Technology}
 %  \streetaddress{P.O. Box 1212}
 	\country{Changsha, China}
 %  \state{Ohio}
 %  \postcode{43017-6221}
}

% \author{Hai Jin}
% %\authornote{School of Computer Science and Technology, HUST, Wuhan, 430074, China}
% \authornotemark[2]
% \authornote{Cluster and Grid Computing Lab, HUST, Wuhan, 430074, China}
% \email{hjin@hust.edu.cn}
% \affiliation{%
% 	\institution{School of Computer Science and Technology, Huazhong University of Science and Technology}
% 	\country{Wuhan, China}
% }

% \renewcommand{\shortauthors}{Shangwen Wang, Ming Wen, Bo Lin, Hongjun Wu, Yihao Qin, Deqing Zou, Xiaoguang Mao, and Hai Jin}

\begin{abstract} 
Retrieval-Augmented Code Generation (RACG) leverages external knowledge to enhance Large Language Models (LLMs) in code synthesis, improving the functional correctness of the generated code. However, existing RACG systems largely overlook security, leading to substantial risks. 
Especially, the poisoning of malicious code into knowledge bases can mislead LLMs, resulting in the generation of insecure outputs, which poses a critical threat in modern software development.
To address this, we propose a security-hardening framework for RACG systems, \toolname, that shifts the paradigm from retrieving only functional code examples to incorporating both functional code and security knowledge.
Our framework constructs a security knowledge base from real-world vulnerability databases, including secure code samples and root cause annotations. For each code generation query, a retriever decomposes the query into fine-grained sub-tasks and fetches relevant security knowledge. To prioritize critical security guidance, we introduce a re-ranking and filtering mechanism by leveraging the LLMs' susceptibility to different vulnerability types. This filtered security knowledge is seamlessly integrated into the generation prompt.
Our evaluation shows \toolname significantly improves code security rates across various LLMs, achieving average improvements of 20.12\% in standard RACG, and 31.53\% and 21.91\% under two distinct poisoning scenarios without compromising functional correctness. Furthermore, \toolname demonstrates strong generalization, enhancing security even when the targeted language's security knowledge is lacking. This work presents \toolname as a pivotal advancement towards building secure and trustworthy RACG systems.
\end{abstract}



% make the title area
\maketitle
\section{Introduction}
\label{sec:intro}

Large Language Models (LLMs) have demonstrated remarkable capabilities across a wide range of domains, from natural language processing to mathematical problem-solving. Their ability to understand and generate human-like text has led to widespread adoption in various applications. Retrieval-augmented generation (RAG) has emerged as a powerful paradigm to further enhance LLMs by leveraging external knowledge bases, enabling more contextually accurate and informed responses. In the domain of code generation, Retrieval-Augmented Code Generation (RACG) has achieved significant advancements by incorporating the relevant knowledge during the code generation (\eg related code snippets), improving the quality of generated code. As a result, RACG-based LLM systems~\cite{microsoft2024,openai2024,su2024evor} have become an indispensable assistant in software development, streamlining the coding process and aiding developers in producing complex software with enhanced accuracy.

Despite these advances, existing RACG systems prioritize functional correctness while often overlooking security considerations. A recent study~\cite{lin2025exploring} highlights critical security vulnerabilities in RACG, particularly when developer intents are exposed to attackers. Specifically, maliciously injected vulnerable code can significantly compromise the security of generated code, with empirical evidence showing that even a single poisoned sample can lead to 48\% of generated code containing vulnerabilities. This threat is further exacerbated in real-world scenarios, where adversaries can introduce a large number of vulnerable samples into the knowledge base, posing a substantial risk to software security. Despite the urgency of this issue, current RACG systems lack dedicated security mechanisms to mitigate such threats.

To mitigate the security threats in RACG, our key idea is to refine the retrieval contents used in the prompt. 
Specifically, our approach revolves around transitioning the workflow from solely retrieving functional code examples to retrieving both functional code examples and security knowledge.
The former aspect ensures that the generated code meets functional requirements, while the latter aspect, which includes secure code samples and annotated root causes of potential vulnerabilities, is dedicated to averting common security vulnerabilities in the generated code. 
That is to say, in addition to providing code snippets for reference implementations, we proactively include security knowledge in the prompts, aiming to assist LLMs in steering clear of vulnerabilities when generating code.
We postulate that incorporating this security knowledge in the prompt can fortify the security defenses of LLMs, enabling them to ``course-correct'' even in scenarios where the knowledge base is poisoned, thereby preventing the generation of vulnerable code.

Building on this intuition, we introduce \toolname, a security-hardening framework for RACG systems designed to achieve both functionality and security in generated code.
\toolname first constructs a security knowledge base by extracting insights (e.g., vulnerability root causes) from historical vulnerabilities.
% Subsequently, for a given code generation query, \toolname employs a context-aware, fine-grained knowledge retriever to identify relevant security knowledge. 
Subsequently, for a given code generation query, \toolname employs an elaborate retriever to identify relevant security knowledge. 
Specifically, \toolname breaks down the query into sub-tasks to retrieve precise security knowledge for each sub-task. Furthermore, recognizing the varying susceptibility of LLMs to different vulnerability types~\cite{tihanyi2025secure}, \toolname re-ranks the retrieved security knowledge based on these susceptibilities, prioritizing knowledge related to more prevalent vulnerabilities and filtering out less relevant knowledge. Finally, the retrieved knowledge is explicitly injected into the prompt as part of the security-augmented code generation process, ensuring that LLMs incorporate security knowledge during code generation while preserving functional correctness.

We rigorously evaluate \toolname across diverse scenarios, including standard RACG and two distinct RACG poisoning scenarios. Our evaluation results demonstrate that \toolname significantly improves the security of generated code, even under poisoning scenarios, without compromising functional correctness across various LLMs and languages. Specifically, in standard RACG scenarios, \toolname achieves an average security improvement of 20.12\%. Under poisoning attacks, \toolname enhances security by 31.53\% and 21.91\% in the respective scenarios.
Furthermore, generalization analysis reveals that \toolname's efficacy extends beyond RACG, improving security in code generation when there is no off-the-shelf knowledge base. Specifically, \toolname achieves an average security rate of 75.54\% across four languages, surpassing the state-of-the-art Safecoder~\cite{he2024instruction} (69.81\%). Notably, even in scenarios where no language-specific secure knowledge is available, \toolname consistently improves security by 15.69\%, 21.26\%, and 13.50\% in standard and poisoning scenarios, respectively.

In summary, our study makes the following contributions:
\begin{itemize}[leftmargin=*]
    \item {\bf Significance:} We introduce the first security-hardening framework for RACG systems, tackling the unaddressed security threats in RACG systems, especially when facing knowledge base poisoning. This work pioneers a critical shift toward securing RACG, a cornerstone of modern LLM-driven software development.

    \item {\bf State-of-the-art Security Hardening Framework:} We propose \toolname, a framework that significantly hardens the security of RACG systems. By explicitly injecting security knowledge into the prompts, \toolname mitigates security risks effectively while maintaining the functionality of generated code.
    
    \item {\bf Extensive Study:} We conduct a rigorous and comprehensive evaluation of \toolname across diverse scenarios, including standard RACG, two RACG poisoning setups, and direct code generation. 
    % Our findings demonstrate that \toolname not only achieves impressive results in RACG scenarios, but also exhibits remarkable generalization capabilities in non-retrieval scenarios.
    Our findings demonstrate that \toolname exhibits remarkable generalization capabilities in various scenarios.
\end{itemize}

\section{Background and Related Works}
\label{sec:bg}

\subsection{Retrieval Augmented Code Generation}
LLMs have seen rapid advancement in recent years, driven by improvements in model architecture, training techniques, and access to large-scale data. Trained on diverse textual sources such as Wikipedia and GitHub, general-purpose models like GPT~\cite{brown2020language} have demonstrated impressive capabilities across a range of tasks, including those in the programming domain.

RAG is a transformative approach that enhances LLMs by incorporating relevant information retrieved from external knowledge sources. This integration significantly improves model performance by combining the strengths of retrieval systems and generative capabilities. RAG has garnered attention across various domains due to its ability to produce more accurate and contextually rich outputs, establishing itself as a robust framework for advancing natural language processing applications.
Building on RAG's success, RACG has emerged as a specialized adaptation tailored to the coding domain. RACG leverages retrieved code snippets, or other programming-related resources to enhance the efficiency and quality of code generation. By integrating domain-specific external knowledge, RACG enables LLMs to tackle complex programming tasks more effectively, surpassing the limitations of purely generative approaches~\cite{zhang2023syntax, gao2024preference,yang2025empirical}. 

\subsection{Security of LLM-Generated Code}
Although LLMs demonstrate significant capabilities in generating functionally correct code, recent studies \cite{pearce2022asleep,wang2024your} highlight persistent difficulties in producing \textit{secure} code. The security implications of LLMs used for code generation have consequently become a major research focus. Pearce~\etal~\cite{pearce2022asleep} systematically evaluated the security of LLM-generated code, focusing on the MITRE Top-25 vulnerabilities. Their findings indicated that approximately 40\% of the generated code contained vulnerabilities, a result corroborated by subsequent studies~\cite{klemmer2024using,tihanyi2025secure}. 
Traditional mitigation involves post-generation scanning using static analysis tools~\cite{codeql, Infer2024}, but this approach introduces latency due to the separate analysis step.

Recently, researchers have proposed approaches to enhance the security of LLM-generated code directly, without relying on retrieval. For example, SafeCoder~\cite{he2024instruction} and SVEN~\cite{he2023large} employ fine-tuning techniques to train LLMs to generate inherently more secure code. CoSec~\cite{li2024cosec} improves security via supervised co-decoding, modifying the generation process without altering the LLM's weights. 
These techniques mainly address non-retrieval scenarios, where LLMs rely only on pretrained knowledge, raising security concerns from training data vulnerabilities. In contrast, RACG systems combine internal knowledge with runtime-retrieved external knowledge, introducing a distinct security challenge from the potential interplay between internal and external knowledge. This challenge has not been sufficiently investigated, and there are no studies specifically addressing security hardening in the context of such interactions.
This gap is critical, especially when the external knowledge base is poisoned with vulnerable code snippets, compromising the generated output. This paper mitigates this gap by introducing \toolname, a security-hardening framework specifically designed for RACG, aimed at fostering the creation of secure and trustworthy code generation systems.

\section{Motivating Examples}
\label{sec:motivation}
\begin{figure}
    \centering
    \begin{subfigure}[b]{\linewidth}
    \centering
        \includegraphics[width=0.85\linewidth]{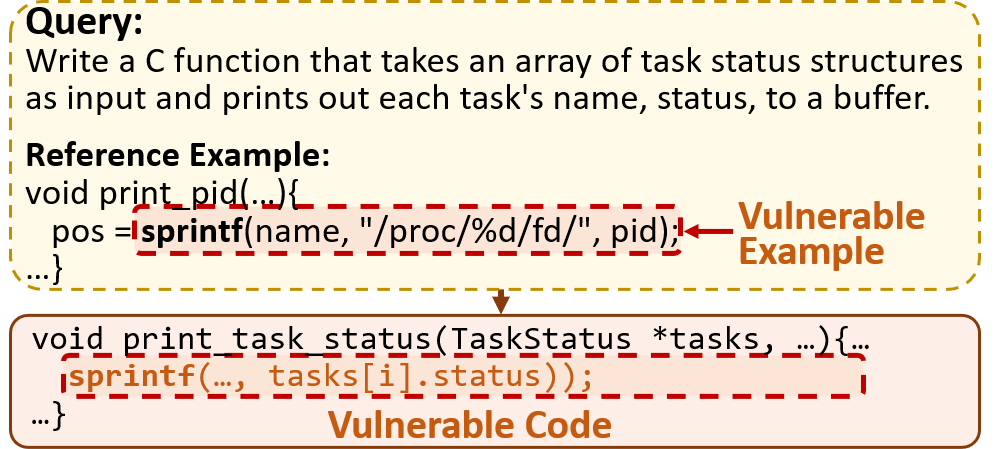}
        \caption{Vulnerable code generated when a poisoned example is retrieved.}
        \label{fig:motivation_a}
    \end{subfigure}
    \begin{subfigure}[b]{\linewidth}
    \centering
        \includegraphics[width=0.85\linewidth]{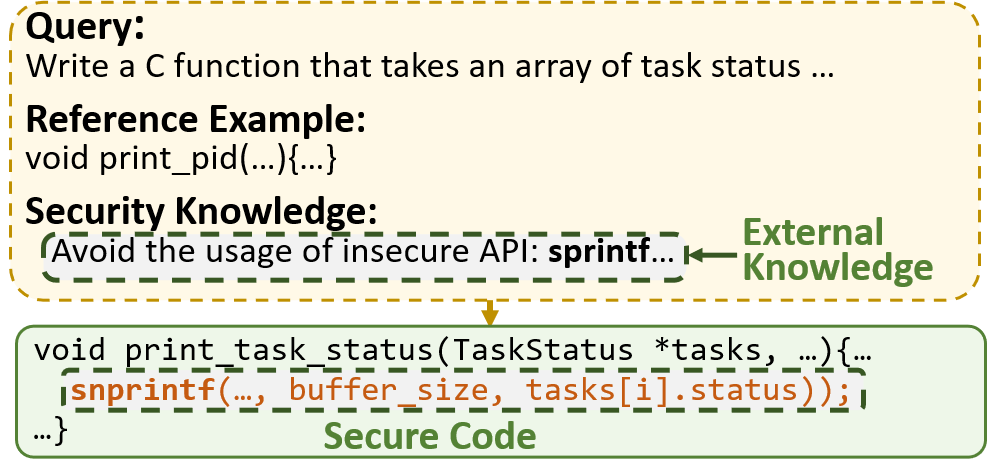} 
        \caption{Secure code generated when augmented with external security knowledge.} 
        \label{fig:motivation_b}
    \end{subfigure}
    \caption{An example of code generated by GPT-4o with and without additional security knowledge.} 
    \label{fig:motivation}
\end{figure}
Existing RACG systems often prioritize functional correctness, potentially overlooking crucial security considerations. This focus can leave them susceptible to generating insecure code, especially in adversarial scenarios. For instance, a prior study~\cite{lin2025exploring} demonstrated that poisoning the retrieval knowledge base with even a single malicious example could lead to vulnerabilities appearing in nearly half (48\%) of the code generated by the LLM.  

Figure~\ref{fig:motivation_a} illustrates such a scenario using a case from CyberSecEval~\cite{bhatt2023purple}, where the code is generated by GPT-4o within a poisoned RACG system. The user's instruction prompts the LLM to generate a function displaying all fields of the given struct. Following standard procedure, the RACG system retrieves a semantically similar code example from its knowledge base. However, in this instance, the knowledge base has been poisoned, and the retrieved example itself contains a vulnerability. 
Specifically, the retrieved example utilizes the {\tt sprintf} function. Guided by this insecure reference, the LLM generates code that also employs {\tt sprintf} to fulfill the user's request. However, {\tt sprintf} is inherently insecure and introduces potential buffer overflow vulnerabilities.

% This observation motivates our core hypothesis: the security of RACG systems can be significantly enhanced by {\bf injecting relevant external security knowledge} into the generation process.
To address the above challenge, our intuition is that if we could {\bf inject relevant external security knowledge} into the prompt, we may help LLMs avoid this mistake. For instance, LLMs may generate secure code if they are reminded that the {\tt sprintf} is a dangerous function. 
Figure~\ref{fig:motivation_b} presents the same code generation task, but this time augmented with such knowledge. Instead of solely relying on retrieved functional code examples, the LLM is provided with relevant security knowledge, which highlights the risks associated with {\tt sprintf} in this case. As a result, when security knowledge is included, the LLM selects the safe alternative {\tt snprintf}, effectively mitigating buffer overflow risks while preserving functionality. 

This example highlights a critical risk in current RACG systems and points towards our proposed solution: hardening RACG systems through the {\bf injection of relevant security knowledge}. Our core idea is to shift the paradigm from solely relying on functional examples towards proactively augmenting the generation with relevant security knowledge. By equipping the LLM with this crucial knowledge, we hypothesize that we can effectively guide it towards generating code that is not only functionally correct but also vulnerability-free, mitigating vulnerabilities like the demonstrated buffer overflow.

\section{Methodology}
\label{sec:methodology}
\begin{figure}
    \centering
    \includegraphics[width=1\linewidth]{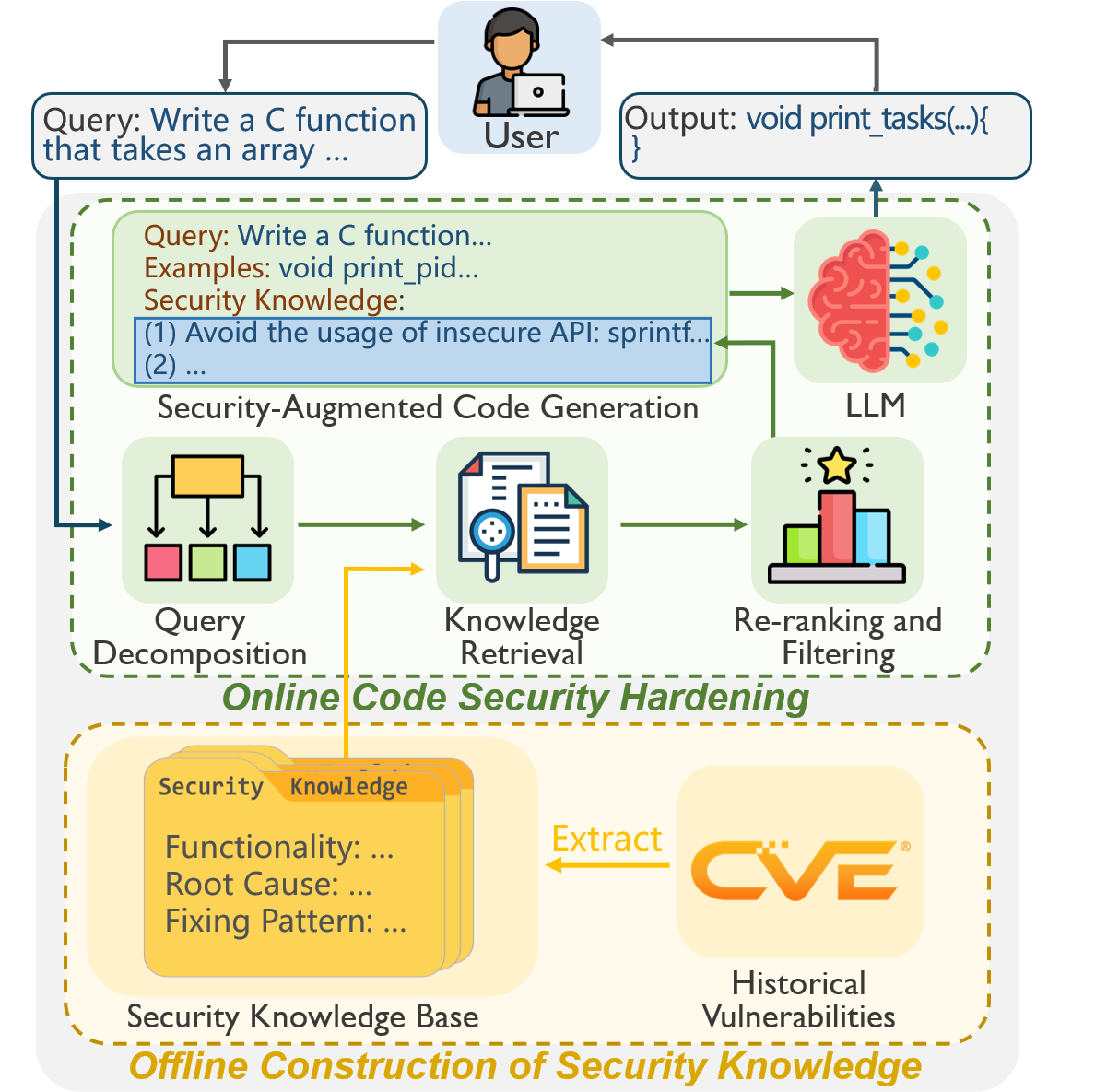}
    \caption{The overall workflow of \toolname}
    \label{fig:framework}
\end{figure}
In this work, we propose \toolname, a security hardening framework for RACG systems that injects security knowledge derived from existing vulnerabilities to harden the security of code generated by RACG systems. 
% As depicted in Figure~\ref{fig:framework}, \toolname enhances the security of generated code by leveraging security knowledge derived from known vulnerabilities. 
As shown in Figure~\ref{fig:framework}, \toolname operates in two primary phases: (1) offline construction of a security knowledge base, and (2) online integration of this knowledge to harden the security of code generated by RACG systems.
In the offline phase, \toolname analyzes historical vulnerabilities to automatically build the security knowledge base (\S\ref{subsec:knowledge_base}). Subsequently, during the online phase, this knowledge base is utilized to harden the code generation process for user queries. This online hardening involves two main stages: first, relevant security knowledge is retrieved based on the input query (\S\ref{subsec:knowledge_retrieval}), and second, this retrieved knowledge is integrated into the prompt to guide the LLM towards generating secure code (\S\ref{subsec:sec_code_gen}). 
% The subsequent sections detail each component of this methodology.}

% Our framework comprises three core phases: (1) offline security knowledge base construction, (2) context-aware fine-grained online knowledge retrieval, and (3) security-augmented code generation. The overall architecture is depicted in Figure~\ref{fig:framework}. Specifically, \toolname begins by constructing a security knowledge base through the analysis of known vulnerabilities using the LLM ( \S\ref{subsec:knowledge_base}). Following this, a context-aware, fine-grained knowledge retriever decomposes code generation instructions into granular sub-tasks, enabling precise extraction of relevant security knowledge (illustrated \S\ref{subsec:knowledge_retrieval}). 
% Next, we re-rank and filter the retrieved security knowledge based on the LLM’s susceptibility to different vulnerability types (\S\ref{subsubsec:reranking}). 
% Finally, in the security-augmented code generation stage, we employ a tailored prompt to integrate these instructions with corresponding security knowledge, constraining the LLM to generate secure code within the RACG system (\S\ref{subsec:sec_code_gen}).

\subsection{Automated Offline Construction of the Security Knowledge Base}
\label{subsec:knowledge_base}
\subsubsection{Security Knowledge Definition}
\label{subsubsec:sec_kng_def}
% \begin{figure}
%     \centering
%     \includegraphics[width=0.9\linewidth]{figures/SecurityKnowledge.png}
%     \caption{An example of security knowledge extracted from vulnerability}
%     \label{fig:security_kng}
% \end{figure}

The security knowledge base, denoted as $\mathcal{S}$, is constructed through an automated offline pipeline that processes historical vulnerabilities from Common Vulnerabilities and Exposures (CVE) instances. Each vulnerability instance $\mathcal{V}_i = (C_v, C_f, D_{cve}, D_{cwe})$ consists of $C_v$ (vulnerable version), $C_f$ (fixed version), $D_{cve}$ (CVE description in natural language), and $D_{cwe}$ (MITRE classification identifier).
\toolname represents the security knowledge in three dimensions: functionality, root cause, and fixing pattern. Figure~\ref{fig:security_kng} (in Appendix) illustrates a representative example extracted from CVE-2012-6538. Each dimension is detailed below:
\begin{itemize}[leftmargin=*]
  \item \textbf{Functionality}: To enable the RACG system to retrieve relevant security knowledge, we construct a representation for each security knowledge. Existing CVE descriptions typically focus on the root cause of vulnerability and impact, whereas the input to the RACG system (i.e., the query) specifies the intended functionality of the code, resulting in a significant semantic gap. To address this, we define the \emph{functionality} associated with security knowledge by describing the functionality of the vulnerable code, extracted using the LLM backend. By leveraging \emph{functionality} to represent corresponding security knowledge, the RACG system can retrieve relevant knowledge effectively.

  % By leveraging \emph{functionality}, the RACG system can retrieve security knowledge more effectively than relying on the similarity between the query and CVE descriptions or between the query and vulnerable code.
  % The \emph{functionality} provides a high-level description of the behavior of the vulnerable code. The \emph{functionality} is served as the representative of the corresponding secure knowledge that is to be retrieved.
  % This description not only helps the LLM understand the semantics of the code but also offers a smaller gap compared to other natural language-based information (e.g., CVE descriptions), which primarily focus on the nature and impact of vulnerabilities. Thus, \emph{functionality} is also representative of the corresponding secure knowledge to be retrieved.
  \item \textbf{Root Cause}: The \emph{root cause} explains why vulnerabilities occur. It consists of two components: (i) a natural language description of the vulnerability's root cause and (ii) a code example illustrating the vulnerability. Together, these provide both conceptual descriptions and detailed code implementations, comprehensively helping the LLM avoid generating vulnerable code.
  \item \textbf{Fixing Pattern}: While the \emph{root cause} helps the LLM recognize vulnerable code, the \emph{fixing pattern} guides the LLM to generate secure code. Similar to the root cause, the fixing pattern consists of two parts: a natural language description of the fix and an example of the corrected code. These provide complementary perspectives on how to resolve vulnerabilities, assisting the LLM in generating secure code.
\end{itemize}

% Figure~\ref{fig:security_kng} (in Appendix) presents an example of security knowledge extracted from CVE-2012-6538. 
% This vulnerability arises from the use of an insecure function, which may lead to a buffer overflow. The patched code mitigates this issue by replacing {\tt strcpy} with {\tt strncpy}, enforcing a specific size constraint on the destination buffer. The \textbf{Functionality} dimension describes the fundamental operation of the vulnerable code snippet (\ie copying a string from a source buffer to a destination buffer). The \textbf{Root Cause} dimension provides a detailed explanation of the vulnerability in natural language along with an illustrative code example (\ie the risks associated with using {\tt strcpy}). Finally, the \textbf{Fixing Pattern} dimension includes both a description of the secure coding practice and an example of the corrected code (\ie replacing {\tt strcpy} with a safer alternative, {\tt strncpy}). This structured knowledge extraction process informs secure code generation in RACG, enabling \toolname to enhance the security of the generated code.

\subsubsection{Security Knowledge Base Construction}
\label{subsubsec:knw_retriever}
For each vulnerability, we prompt the LLM to extract the security knowledge. Prompt~\ref{prompt:1} (in Appendix) provides the template used for the extraction. 
This template takes as input the CVE description ($D_{\text{cve}}$), the CWE classification type ($D_{\text{cwe}}$), and the function-level diff (\texttt{DIFF}), which compares the vulnerable code ($C_v$) with its patched version ($C_f$).
% The terms $D_{\text{cve}}$ and $D_{\text{cwe}}$ denote the CVE description and the Common Weakness Enumeration (CWE) classification type, respectively. The \texttt{DIFF} represents the function-level diff, generated by comparing the vulnerable code $C_v$ with its corresponding fixed version $C_f$. Note that \texttt{DIFF} includes the full function context, as this detailed diff provides comprehensive information for understanding both the vulnerability and its resolution.
% \input{prompts/KnowExtract}
The output for each vulnerability is a tuple consisting of a functionality description, root cause, and fixing pattern. This process is repeated for all instances in the dataset, and the extracted items are aggregated to form the security knowledge base $\mathcal{S}$.

\subsection{Context-Aware Fine-Grained Knowledge Online Retrieval}
\label{subsec:knowledge_retrieval}
In this stage, we employ a context-aware, fine-grained knowledge retriever to dynamically extract relevant knowledge $\mathcal{S}_Q$ for a given query $Q$ from the constructed security knowledge base $\mathcal{S}$. This process comprises the following key stages: query decomposition, security knowledge retrieval, re-ranking, and filtering.

\subsubsection{Query Decomposition}
\label{subsubsec:query_decomp}
\begin{figure*}
    \centering
    \includegraphics[width=0.85\linewidth]{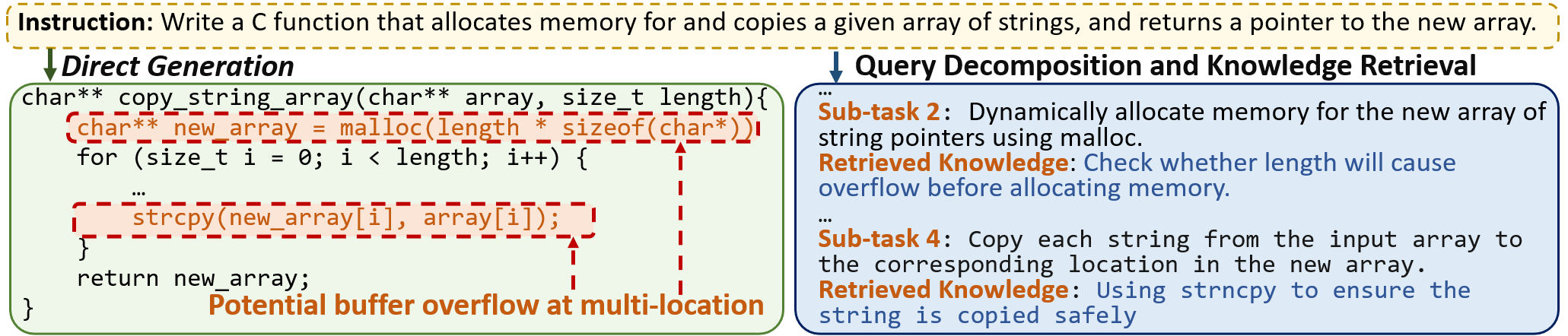}
    \caption{Vulnerable code generated by GPT-4o along with its corresponding decomposed queries and retrieved knowledge.}
    \label{fig:query_decomp}
\end{figure*}
As suggested in previous studies~\cite{huang2024vmud, he2025vultr}, a single program may contain multiple vulnerabilities across different locations, meaning that a code generated by a single query may correspond to multiple security issues. 
% This observation motivates us to decompose the user's query into finer semantic units, enabling the retrieval of more relevant and specific security knowledge, thereby enhancing the security of generated code.
Figure~\ref{fig:query_decomp} presents a concrete example from CyberSecEval~\cite{wan2024cyberseceval}, generated by GPT-4o. The query instructs the LLM to generate a function that copies a string and returns the corresponding pointer. However, the generated code introduces multiple buffer overflow vulnerabilities related to memory allocation and string copying at different locations. This insight led us to decompose the user’s query into smaller, fine-grained semantic units (\ie sub-tasks), enabling more precise knowledge retrieval and enhancing the security of the generated code, as illustrated in Figure~\ref{fig:query_decomp}.
For example, for the decomposed sub-task related to memory allocation, the retrieved knowledge explicitly indicates that the length should be checked to prevent overflow before allocating memory. 
% Further details on knowledge retrieval are provided in \S\ref{subsubsec:know_retrieval}.

LLMs have demonstrated a remarkable capability in understanding complex natural language instructions and discerning underlying semantic components. Their ability to process intricate queries and apply reasoning makes them well-suited for breaking down tasks into concrete steps, as demonstrated by techniques like chain-of-thought~\cite{wei2022chain,li2025structured}.
Leveraging this capability, we employ the LLM to decompose the query into smaller sub-tasks that represent the query’s semantics at a finer level. The decomposition template is detailed in Prompt~\ref{prompt:2} (see Appendix). The final decomposed query $Q_{d}$ consists of a list of sub-tasks $Q_{d} = [q_1, q_1, \dots, q_n]$, which are then used in the knowledge retrieval stage.

\subsubsection{Security Knowledge Retrieval}
\label{subsubsec:know_retrieval}
In this stage, we aim to preliminarily identify the relevant security knowledge entries from the security knowledge base $\mathcal{S}$. We adopt a similarity-based retrieval approach, using cosine similarity to measure the relevance between the feature representations of the sub-tasks and the knowledge entries in the base.
Formally, for a list of decomposed sub-tasks \( Q_{d} = [q_1, q_1, \dots, q_n] \), where each \( q_i \) represents a fine-grained semantic unit (\ie sub-task), we compute the feature vector for each sub-task \( q_i \) by embedding its description into a vector:
\[
\mathbf{v}_{q_i} = \text{Embed}(q_i),
\]
where \( \mathbf{v}_{q_i} \in \mathbb{R}^d \) is the feature vector for sub-task \( q_i \), and \(\text{Embed}(\cdot)\) denotes the embedding function (detailed in \S\ref{subsec:retriever}).
Each knowledge entry \( s_j \) in $\mathcal{S}$ is also represented by a feature vector \( \mathbf{v}_{s_j} \), computed similarly by embedding its textual description:
\[
\mathbf{v}_{s_j} = \text{Embed}(\text{desc}(s_j)),
\]
where \( \text{desc}(s_j) \) is the textual description of the functionality of the knowledge entry \( s_j \) (as defined in \S\ref{subsubsec:sec_kng_def}).

To retrieve the relevant knowledge for a given sub-task \( q_i \), we calculate the cosine similarity between its feature vector \( \mathbf{v}_{q_i} \) and the feature vector \( \mathbf{v}_{s_j} \) of each knowledge entry \( s_j \) in the knowledge base. 
Based on these similarity scores, we retrieve the top-$k'$ (discussed in \S\ref{subsec:num_kng}) most relevant knowledge entries for $q_i$, denoted as $\mathcal{S}'_{q_i}$, which are then subjected to subsequent re-ranking and filtering.
% we combine these results into a preliminary knowledge base \( \mathcal{S'}_Q \) for the original query \( Q \):
% \[
% \mathcal{S'}_Q = \bigcup_{i=1}^{n} \mathcal{S}'_{q_i}.
% \]

\subsubsection{Security Knowledge Re-ranking and Filtering}
\label{subsubsec:reranking}
\begin{table}[!t]
  \centering
  \caption{Distribution of violation types and associated CWEs in LLM-generated code}
  \resizebox{0.90\linewidth}{!}{
  \begin{tabular}{llll}
    \toprule
    {\bf Violation Type} & {\bf Related CWEs} & {\bf Percentage} \\
    \midrule
\rowcolor{lightgray} NULL pointer & 391, 476, 690; & 40.24\% \\
    Buffer overflow & 120, 121, 122, 628, 676 & 25.53\%\\
    & 680, 787; \\
\rowcolor{lightgray} Invalid pointer & 822, 119; & 10.42\% \\
    Array bounds violated & 125, 129, 131, 193, 788; & 8.86\% \\
\rowcolor{lightgray} Arithmetic overflow & 191, 20, 190, 192, 681;& 6.21\%\\
 Resource mismanagement & 825, 401, 404, 459; & 5.03\% \\
 \rowcolor{lightgray}   Division by zero & 369, 691; & 1.45\% \\
    Others & - & 2.26\% \\
    \bottomrule
  \end{tabular}
  }
  \label{tab:violation_types}
\end{table}
Note that before this stage, we have retrieved relevant security knowledge for each sub-task. Although the retrieval stage ensures that each sub-task is associated with a fixed number of top-$k'$ security knowledge entries, the effectiveness of this approach is limited by two key challenges. First, the input window size of the LLM may be exceeded, leading to truncation or loss of critical context~\cite{liu2024lost,chang2024efficient}. Second, LLMs often struggle with long, complex prompts due to attention dilution, reducing their ability to leverage external knowledge~\cite{zhao2023survey,xiao2023efficient}.
  
To mitigate these challenges and refine the preliminary knowledge obtained during the retrieval stage, we introduce a re-ranking and filtering mechanism that tailors the security knowledge to the varying risk profiles of individual sub-tasks. The key insight is that not all sub-tasks exhibit the same susceptibility to vulnerabilities. For example, a sub-task responsible for outputting results typically poses less risk than one involving memory allocation, which is more prone to issues like buffer overflows. 
To quantify the risk associated with each sub-task $q_i$, we first determine the weight $w_j$ for each individual knowledge entry $s_j$ within the initially retrieved set $\mathcal{S}'_{q_i}$. This weight reflects the likelihood that the vulnerability associated with $s_j$ manifests in LLM-generated code.
% To address this disparity and reduce the interference of extraneous information on the LLM, we re-rank the retrieved security knowledge according to the risk level collected offline. Specifically, we assess the risk level of a sub-task by the likelihood that vulnerabilities associated with its retrieved security knowledge manifest in LLM-generated code. 
This approach is grounded in the observation that sub-tasks semantically related to frequently occurring vulnerabilities are more likely to induce the LLM to produce insecure code. To quantify this likelihood, we draw on a prior study~\cite{tihanyi2025secure}, which analyzed the distribution of violation types and their associated CWEs across code generated by 13 mainstream LLMs (\eg GPT and Gemini series models) on the dataset containing 310,531 code generation instructions. The resulting distribution, detailed in Table~\ref{tab:violation_types}, provides empirical evidence of vulnerability prevalence in LLM-generated code.

For each knowledge entry $s_j$ within the preliminary retrieved set $\mathcal{S}'_{q_i}$ for sub-task $q_i$, we assign a weight $w_j$ reflecting the probability of its associated vulnerability appearing in LLM-generated code. Specifically, we map the CWE of $s_j$, to its corresponding violation type in Table~\ref{tab:violation_types}. The weight $w_j$ is then set to the percentage frequency of that violation type. For example, a knowledge entry related to CWE-476 (NULL pointer dereference), categorized under the "NULL pointer" violation type with a frequency of 40.24\%, receives a weight $w_j = 0.40$. For CWEs not listed in the table, we assign a default weight of $w_j = 0.01$ to ensure all entries are preliminarily considered.
Next, we calculate the overall weight $W_{q_i}$ for each sub-task $q_i$ by summing the weights of all its initially retrieved knowledge entries:
\[
W_{q_i} = \sum_{s_j \in \mathcal{S}'_{q_i}} w_j.
\]
This aggregated weight $W_{q_i}$ represents the estimated risk level of sub-task $q_i$, considering the vulnerabilities associated with its relevant knowledge.

We then re-rank all sub-tasks $[q_1, ..., q_n]$ based on weights $W_{q_i}$ in descending order. Subsequently, we perform a filtering step by selecting the sub-tasks corresponding to the top-$k$ highest weights. Let $\mathcal{Q}_{\text{top-}k}$ denote this set of top-$k$ sub-tasks. The parameter $k$ is set to five as discussed in  \S\ref{subsec:num_kng}. This filtering strategy ensures that the security knowledge associated with the sub-tasks deemed most likely to introduce critical vulnerabilities is retained. By prioritizing sub-tasks with high aggregated weights, this process naturally emphasizes knowledge tied to high-risk vulnerabilities.
% like buffer overflows or NULL pointer dereferences.
 % implicitly focusing the guidance provided to the LLM.}

The final security knowledge base $\mathcal{S}_Q$ for the query $Q$ is then constructed by aggregating all the initially retrieved knowledge entries corresponding to these selected top-$k$ sub-tasks:
\[
\mathcal{S}_Q = \bigcup_{q_i \in \mathcal{Q}_{\text{top-}k}} \mathcal{S}'_{q_i}.
\]
This curated $\mathcal{S}_Q$ is subsequently integrated into the code generation process (see \S\ref{subsec:sec_code_gen}), enabling \toolname to guide the LLM toward producing secure code by concentrating on the highest-risk aspects identified through sub-task ranking.

\subsection{Security-Augmented Code Generation}
\label{subsec:sec_code_gen}
In conventional RACG systems, LLMs generate code by leveraging both a user-provided query and external knowledge retrieved from the knowledge base, typically, code snippets. The prompt structure in traditional RACG can be expressed as:
\[
P_{ori} = Q + E,
\]
where \(Q\) is the user’s input specifying the desired functionality, and \(E\) provides example implementations.
% For example, if a user requests a function to allocate memory for a string and copy it, the retrieved knowledge might include a secure implementation using {\tt malloc} and {\tt strcpy}. This augmentation improves the LLM’s ability to produce functionally correct code by grounding its output in real-world examples or best practices. However, traditional RACG often neglects security. The generated code may harbor vulnerabilities, stemming from inherent limitations in LLMs or external factors, such as a poisoned knowledge base introducing insecure examples.

To enhance the security of RACG systems, \toolname injects security knowledge \( \mathcal{S}_Q \) into the code generation prompt. Specifically, for a given user query \( Q \) and its filtered sub-tasks \( q_i \in \mathcal{Q}_{\text{top-}k} \), each sub-task \( q_i \) is paired with its corresponding security knowledge \( \mathcal{S}'_{q_i} \). The security-augmented prompt is constructed as:
\[
P = P_{ori} + \sum_{q_i \in \mathcal{Q}_{\text{top-}k}} \left( q_i + \mathcal{S}'_{q_i} \right),
\]
which ensures that LLM generates code satisfying both functional requirements and security constraints. The Prompt~\ref{prompt:3} (in Appendix) provides an example of this process.
For instance, if \(q_i\) involves ``copy a string from input to destination'', \(\mathcal{S}'_{q_i}\) might include guidelines like ``Use \texttt{strncpy} to ensure the string is copied safely''. 
% If no security knowledge is retrieved for a sub-task (i.e., filtered out by re-ranking in \S\ref{subsubsec:reranking}), the corresponding field remains blank.

% \input{prompts/CodeGen}

\section{Study Design}
\label{sec:setup}

\subsection{Investigated Scenarios}
\label{subsec:scenarios}
In this study, we comprehensively evaluate the effectiveness of \toolname across three scenarios: a standard RACG and two poisoning scenarios, following prior work~\cite{lin2025exploring}. The standard scenario represents a typical RACG setting, where the retriever fetches knowledge from a knowledge base free from vulnerable code. In contrast, the two poisoning scenarios simulate situations where programming intents (\ie queries) are either exposed or not exposed to an attacker, who subsequently injects vulnerable code into the knowledge base.

\subsubsection{Standard RACG Scenario}
\label{subsubsec:standard}
In this scenario, the RACG system operates without poisoning. The system retrieves code from the functional code base $\mathcal{K}$ based on the query $Q$ and generates code accordingly. This setting evaluates the framework’s performance in a typical scenario, serving as a reference for the poisoning scenarios.

\subsubsection{Poisoning Scenario I: Exposed Programming Intents}
\label{subsubsec:scenario_1}
In poisoning scenarios, we assume that the attacker can access and poison $\mathcal{K}$, which is typically sourced from public repositories such as GitHub. However, access to programming intents (\ie query $Q$) depends on the attacker's capabilities and may not always be feasible.
In this scenario, we assume the attacker has access to $Q$ and exploits this information to inject vulnerabilities into $\mathcal{K}$. Specifically, the attacker selects the $m$ most semantically similar vulnerable examples from the vulnerable code base $\mathcal{V}$ using a poisoning retriever. The knowledge base used by RACG then becomes $\mathcal{K} \cup \mathcal{V}$, blending secure and vulnerable code. This scenario tests the \toolname's ability to mitigate targeted attacks leveraging intent-specific vulnerabilities. Prior empirical studies~\cite{lin2025exploring} indicate that varying the poisoning quantity exhibits similar patterns in its impact on the security and functionality of the generated code. For clarity, we assess \toolname under a moderate poisoning level, using five poisoned samples (\ie $m = 5$) for each query.

\subsubsection{Poisoning Scenario II: Intent-Agnostic Poisoning}
\label{subsubsec:scenario_2}
In this scenario, we mimic a stricter attack setting than scenario I, where the attacker lacks direct access to $Q$ and instead poisons $\mathcal{K}$ with broadly representative vulnerable code likely to be retrieved across diverse queries. Specifically, prior work~\cite{lin2025exploring} suggests that an attacker can poison the knowledge base by injecting common functionalities that are more likely to be retrieved in RACG, thereby affecting a broader range of queries. Following this insight, we adopt a clustering-based approach to select the top $p$\% most representative examples from $\mathcal{K}$. For each selected example, we retrieve its most similar vulnerable code from $\mathcal{V}$ and inject it into the knowledge base. The resulting $\mathcal{K} \cup \mathcal{V}$ simulates a generalized poisoning attack, evaluating \toolname’s resilience when programming intents are not exposed. In this scenario, we assess \toolname under a moderate poisoning level, using the poisoning proportion of 10\% (\ie $p = 10$). 

% \vspace{-1mm}
\subsection{Benchmark and Knowledge Bases}
\label{subse:benchmark}
\subsubsection{Evaluation Benchmark}

To evaluate the security and functionality of LLM-generated code across diverse scenarios, we adopt CyberSecEval~\cite{wan2024cyberseceval} as our benchmark for two primary reasons:  
(1) It serves as the official benchmark for the LLaMA series of LLMs, which are widely utilized in academia and industry~\cite{kavian2024llm,hariharan2024rethinking}.
(2) It provides the most comprehensive assessment of security among available benchmarks.
Specifically, CyberSecEval comprises 1,916 instances spanning 50 CWE types, exhibiting a lead over the second largest security evaluation benchmark~\cite{tony2023llmseceval}, which includes only 150 instances across 18 CWE types. Additionally, CyberSecEval includes a specialized insecure code detector, built upon analysis rules tailored to its cases, enabling security evaluation with a precision of 96\%.
% This scale and dedicated evaluator make CyberSecEval an ideal choice for rigorously assessing the security of generated code.
% Note that while some datasets, such as ReposVul~\cite{wang2024reposvul}, cover more vulnerabilities, they focus solely on collecting vulnerabilities without providing a corresponding evaluation framework.
While other datasets like ReposVul~\cite{wang2024reposvul} contain more vulnerabilities, they primarily function as repositories of vulnerabilities and lack the associated evaluation framework (e.g., analysis rules) necessary for performing high-precision security assessments on generated code. Consequently, datasets like ReposVul are less suited for rigorously evaluating the security of generated code.
% though potentially valuable for knowledge extraction.

% Note that for RQ1 and RQ2, we use only C, C++, Java, and Python instances from CyberSecEval, as we extract security knowledge from ReposVul~\cite{wang2024reposvul}, which includes these four languages. In RQ3, we assess \toolname’s effectiveness when security knowledge for the target language is unavailable by leveraging CyberSecEval instances from four additional languages: C\#, JavaScript, PHP, and Rust.

\subsubsection{Knowledge Bases Construction}
As illustrated in \S\ref{subsec:scenarios}, the three scenarios involve distinct knowledge bases that serve different roles in the RACG system. 
The first is a security knowledge base that stores security-related knowledge, which \toolname leverages to enhance the security of generated code.  
The second is a knowledge base from which the RACG system retrieves code examples for code generation. The third is a set of vulnerable code, which serves as the attacker's resource for selecting and injecting malicious code into the RACG system, thereby compromising the security of the generated code.
Therefore, three distinct knowledge bases are required: the Security Knowledge Base ($\mathcal{S}$), the Functional Code Base ($\mathcal{K}$) and the Vulnerable Code Base ($\mathcal{V}$), as detailed below:

\begin{itemize}[leftmargin=*]
    \item \textbf{Security Knowledge Base ($\mathcal{S}$):} This base provides security knowledge to harden the security of generated code. It is constructed by analyzing vulnerabilities and their corresponding fixes from ReposVul~\cite{wang2024reposvul}, the largest cross-language dataset of real-world vulnerabilities, which includes 12,053 function-level pairs of vulnerable and secure code across four programming languages. The construction process is described in detail in \S\ref{subsec:knowledge_base}.
    
    \item \textbf{Functional Code Base ($\mathcal{K}$):} This forms the foundational knowledge base for retrieval-augmentation in the RACG system, enabling it to retrieve relevant functional and secure code based on user queries. Commonly used code bases (e.g., CSN~\cite{husain2019codesearchnet}) are typically collected from open source projects, potentially introducing vulnerable code into the dataset. For example, in our analysis of CSN, 81.3\% (684 out of 841) of Java code invoking random-related functions rely on weak randomness, which can lead to serious security vulnerabilities. The presence of such insecure code in functional code bases may bias evaluations and obscure the actual impact of poisoning attacks. To mitigate this, we construct $\mathcal{K}$ using the fixed code (\ie the vulnerability-free code) from the ReposVul dataset, ensuring that all retrieved examples are functionally correct and security-vetted.
    
    \item \textbf{Vulnerable Code Base ($\mathcal{V}$):} To simulate realistic RACG knowledge base poisoning attacks, a dedicated vulnerable code base is essential. This base serves as the source of vulnerable code that attackers would retrieve and inject into the functional code base $\mathcal{K}$ during poisoning scenarios (detailed in \S\ref{subsec:scenarios}). Importantly, $\mathcal{V}$ is built using vulnerable code instances from the CyberSecEval dataset, rather than ReposVul. Using the same dataset for both the vulnerable code and the security knowledge base could lead to overlapping fix strategies, making poisoning attempts easier to detect and defend against, thereby undermining the realism of the poisoning simulation. It is worth noting that this experimental setup grants attackers slightly more power than in real-world scenarios, as they can inject vulnerable code into $\mathcal{K}$ that closely matches the user’s query. This assumption enables us to more rigorously evaluate the effectiveness of \toolname under more severe security threats.

\end{itemize}
% Table generated by Excel2LaTeX from sheet 'Other'
\begin{table}[!t]
  \centering
  \caption{Statistics of knowledge bases}
  \resizebox{0.85\linewidth}{!}{
   \begin{tabular}{lrrrr}
    \toprule
    \multicolumn{1}{c}{\multirow{2}[2]{*}{\bf Knowledge Base}} & \multicolumn{4}{c}{\bf Language} \\
          & \multicolumn{1}{l}{\bf C} & \multicolumn{1}{l}{\bf C++} & \multicolumn{1}{l}{\bf Java} & \multicolumn{1}{l}{\bf Python} \\
    \midrule
    Functional Code Base & 6,956 & 510   & 2,810 & 1,777 \\
    Vulnerable Code Base  & 227   & 259   & 235   & 229 \\
    Security Knowledge Base & 8,861 & 644   & 3,365 & 2,217 \\
    \bottomrule
    \end{tabular}%
  }
  \label{tab:sta_KG}%
\end{table}%

Table~\ref{tab:sta_KG} presents the statistics of the aforementioned knowledge bases across different programming languages. The number of security knowledge instances exceeds the number of vulnerabilities in ReposVul, as some vulnerabilities have multiple root causes, as observed in previous studies~\cite{huang2024vmud, he2025vultr}.

% \subsection{Result Validation}
% \label{subsec:result_validation}
% Our evaluation centers on the security of code generated by the RACG systems, validated using the insecure code detector from CyberSecEval~\cite{wan2024cyberseceval}. This static analysis tool detects vulnerable code across seven programming languages and over 50 CWE types, achieving a precision of 96\% and recall 76\% of LLM-generated code~\cite{wan2024cyberseceval}. We apply it to assess the prevalence of vulnerabilities in code produced under each scenario, comparing outcomes against baseline RACG.

\subsection{Studied LLMs}
\begin{table}[!t]
  \centering
  \caption{Studied LLMs in the study}
  \resizebox{1\linewidth}{!}{
  \begin{tabular}{cccc}
    \toprule
    \textbf{Category} & \textbf{LLM} & \textbf{Publisher} & \textbf{Open-source} \\
    \midrule
          \multicolumn{1}{c}{\multirow{2}[1]{*}{General}} & GPT-4o~\cite{openai_gpt4o} & OpenAI & No \\
    &DeepSeek-V3~\cite{liu2024deepseek} & DeepSeek & Yes \\
    \midrule
        \multicolumn{1}{c}{\multirow{2}[1]{*}{Code}}  & CodeLlama-13B~\cite{roziere2023code} & Meta & Yes \\
     & DeepSeek-Coder-V2-16B~\cite{zhu2024deepseekcoder} & DeepSeek & Yes \\
    \bottomrule
  \end{tabular}
  }
  \label{tab:llms}
\end{table}
To evaluate \textit{\toolname}’s effectiveness, we employ four representative LLMs, encompassing a range of model sizes, and categories. These include: (1) GPT-4o, a closed-source general-purpose model accessed via its API; and (2) DeepSeek-V3, a state-of-the-art general LLM, also accessed via API~\cite{openai2024api}; (3) CodeLlama-13B, a code-oriented open-source model; (4) DeepSeek-Coder-V2-16B, an advanced code-oriented open-source model.  Table~\ref{tab:llms} summarizes their key attributes, spanning small-scale (e.g., 8B–16B) to large-scale models and both general-purpose and code-specific designs. Model weights for open-source LLMs (CodeLlama and DeepSeek-Coder-V2) were sourced from their official Hugging Face repositories. For brevity, we denote these as GPT-4o, DS-V3, CodeLlama and DS-Coder in subsequent sections. 

\subsection{Retriever}
\label{subsec:retriever}
Retrievers are integral to the RACG system, enabling the retrieval of additional knowledge to enhance generation. In this study, we involve three types of retrievers: code retriever, knowledge retriever, and poisoning retriever as follows:
\begin{itemize}[leftmargin=*]
    \item {\bf Code Retriever.} The code retriever supports the RACG pipeline by fetching relevant code examples from the functional knowledge base $\mathcal{K}$ to serve as references during generation. We implement this retriever using the state-of-the-art jina-embeddings-v3 model~\cite{sturua2024jina}, a dense retriever that embeds code snippets into a vector space for similarity-based retrieval. 
    
    \item {\bf Knowledge Retriever.} The knowledge retriever, detailed in \S\ref{subsec:knowledge_retrieval}, underpins our context-aware fine-grained knowledge online retrieval process. It extracts security knowledge from $\mathcal{S}$ by calculating the similarity between sub-tasks (decomposed from the query) and the knowledge entry in the security knowledge base. Note that we utilize the same embedding model for code and knowledge retriever (\ie \texttt{jina-embeddings-v3}) to reduce the semantic gap between the retrieved code and secure knowledge.

    \item {\bf Poisoning Retriever.} In adversarial scenarios, attackers lack access to the parameters or query capabilities of the retrievers in the RACG system. Thus, an external poisoning retriever is required for the retrieval of vulnerable code in poisoning scenario I and generating embeddings for clustering-based poisoning in poisoning scenario II. Therefore, we built he poisoning retriever on the \texttt{text-embedding-3-large} model~\cite{OpenAI_Embeddings} following a previous study~\cite{lin2025exploring}. Operating independently of the RACG system, the poisoning retriever embeds and retrieves vulnerable code from the vulnerable knowledge base to poison the RACG system, simulating the realistic poisoning process of the RACG system.
    
\end{itemize}

\subsection{Metrics}
To evaluate the effectiveness of \toolname in enhancing security while maintaining functionality, we employ the following metrics:

{\bf Security Rate (SR)}: SR quantifies the likelihood of an LLM generating secure code, defined as the percentage of generated code verified as secure. The verification process is conducted using the Insecure Code Detector from CyberSecEval~\cite{wan2024cyberseceval}, which detects vulnerable code across seven programming languages and over 50 CWE types with a precision of 96\%~\cite{wan2024cyberseceval}.

{\bf Similarity (Sim)}: Due to the absence of test cases in the CyberSecEval dataset, we cannot directly evaluate whether \toolname affects the functionality of LLM-generated code. Instead, we use CodeBLEU~\cite{ren2020codebleu}, a BLEU variant for code similarity, to compare generated code with ground truth, as a partial representative for functionality.
Besides, we further assess the functional correctness of LLM-generated code after applying \toolname using test-based benchmarks: MBPP\cite{austin2021program} and HumanEval\cite{chen2021evaluating}, as detailed in \S\ref{subsec:functionality_eval}.

\subsection{Implementation Details}
\label{subsec:implement_details}
All experiments were conducted on an A100 GPU server using the Ollama~\cite{ollama_website}. To ensure output consistency across LLMs, we set the temperature parameter to 0, minimizing non-determinism as recommended by prior work~\cite{ouyang2024empirical}. Model configurations followed established settings~\cite{lin2025exploring}: a \texttt{max\_new\_tokens} limit of 4096, and a context window of 8192 tokens, with other parameters left at defaults. 
% \update{For the selection of knowledge entries in each sub-task ($k'$ in \S\ref{subsubsec:know_retrieval}) and the final number of retrieved knowledge entries ($k$ in \S\ref{subsubsec:reranking}), we carefully considered the length of knowledge entries and the context window size of large models. Based on these factors, we set $k' = 2$ and $k = 5$.}  
We adhered to each model’s recommended prompt formats, sourced from official documentation, GitHub repositories, or original papers, using predefined chat templates where applicable. For constructing the security knowledge base (\S\ref{subsec:knowledge_base}) and performing query decomposition (\S\ref{subsubsec:query_decomp}), we leveraged DeepSeek-V3~\cite{liu2024deepseek}, a state-of-the-art open-source LLM, as the backend. 

\section{Evaluation}
\label{sec:eval}

\subsection{Research Questions}
To systematically evaluate the effectiveness of \toolname in enhancing the security of RACG systems, we formulate the following research questions (RQs):

{\bf RQ1: Performance in Standard RACG.} How does \toolname perform in a non-poisoned RACG scenario?

{\bf RQ2: Resilience to Poisoning Attacks.} How resilient is \toolname to poisoning attacks in RACG scenario?

{\bf RQ3: Generalization of \toolname.}  
How well does \toolname generalize across diverse conditions, including code generation when there is no off-the-shelf knowledge base, and scenarios lacking target-language security knowledge?

\subsection{RQ1: Performance in the Standard RACG}
This RQ evaluates \toolname's ability to enhance the security of RACG-generated code under the standard setting, where the knowledge base comprises solely functional code examples without poisoning. The goal is to assess whether \toolname can effectively leverage retrieved security knowledge to mitigate vulnerabilities.
\begin{table}[!t]
  \centering
  \caption{Performance of \toolname under Standard RACG Scenario.}
  \resizebox{1\linewidth}{!}{
    \begin{tabular}{ccrrrrr} % 
      \toprule
      \multirow{2}[2]{*}{\bf Metrics} & \multirow{2}[2]{*}{\bf LLM} & \multicolumn{5}{c}{\bf Language} \\ 
      \cmidrule{3-7} & & \multicolumn{1}{r}{\bf C} & \multicolumn{1}{r}{\bf C++} & \multicolumn{1}{r}{\bf Java} & \multicolumn{1}{r}{\bf Python} & \multicolumn{1}{r}{\bf Average} \\
      \midrule
      \multicolumn{1}{c}{\multirow{14}[2]{*}{SR}} & {\bf GPT-4o} & 55.95 & 80.31 & 41.92 & 70.09 & 62.07 \\ 
      & w. \toolname$\dagger$ & 70.04 & 84.94 & 62.45 & 81.20 & 74.66 \\ 
      &    & ($\uparrow$ {\bf 25.18\%}) & ($\uparrow$ {\bf 5.77\%}) & ($\uparrow$ {\bf 48.97\%}) & ($\uparrow$ {\bf 15.85\%}) & ($\uparrow$ {\bf 20.28\%}) \\
      & {\bf DS-V3} & 55.07 & 74.90 & 40.61 & 68.95 & 59.88 \\ 
      & w. \toolname & 72.25 & 88.03 & 73.36 & 79.49 & 78.28 \\ 
      &  & ($\uparrow$ {\bf 31.20\%}) & ($\uparrow$ {\bf 17.53\%}) & ($\uparrow$ {\bf 80.65\%}) & ($\uparrow$ {\bf 15.29\%}) & ($\uparrow$ {\bf 30.73\%}) \\ 
      & {\bf CodeLlama} & 52.42 & 74.13 & 44.98 & 68.66 & 60.05 \\ 
      & w. \toolname & 66.52 & 80.31 & 57.21 & 74.93 & 69.74 \\ 
      & & ($\uparrow$ {\bf 26.90\%}) & ($\uparrow$ {\bf 8.34\%}) & ($\uparrow$ {\bf 27.19\%}) & ($\uparrow$ {\bf 9.13\%}) & ($\uparrow$ {\bf 16.15\%}) \\
      & {\bf DS-Coder} & 51.98 & 75.68 & 46.29 & 71.51 & 61.37 \\ 
      & w. \toolname & 65.20 & 80.31 & 55.02 & 78.06 & 69.65 \\ 
      & & ($\uparrow$ {\bf 25.43\%}) & ($\uparrow$ {\bf 6.12\%}) & ($\uparrow$ {\bf 18.86\%}) & ($\uparrow$ {\bf 9.16\%}) & ($\uparrow$ {\bf 13.50\%}) \\ 
      \cmidrule{2-7}
      & Average & 53.86 & 76.26 & 43.45 & 69.80 & 60.84 \\ 
      & w. \toolname & 68.50 & 83.40 & 62.01 & 78.42 & 73.08 \\ 
      & & ($\uparrow$ {\bf 27.20\%}) & ($\uparrow$ {\bf 9.37\%}) & ($\uparrow$ {\bf 47.72\%}) & ($\uparrow$ {\bf 12.35\%}) & ($\uparrow$ {\bf 20.12\%}) \\ 
      \midrule
      \multicolumn{1}{c}{\multirow{10}[4]{*}{Sim}} & {\bf GPT-4o} & 22.82 & 24.51 & 28.56 & 21.31 & 24.30 \\ 
      & w. \toolname & 23.72 & 25.45 & 29.17 & 23.36 & 25.43 \\ 
      & {\bf DS-V3} & 22.76 & 24.10 & 28.98 & 21.10 & 24.24 \\ 
      & w. \toolname & 24.11 & 25.36 & 29.31 & 23.46 & 25.56 \\ 
      & {\bf CodeLlama} & 23.14 & 24.31 & 28.01 & 21.73 & 24.30 \\ 
      & w. \toolname & 23.89 & 25.65 & 28.34 & 22.95 & 25.21 \\ 
      & {\bf DS-Coder} & 21.95 & 24.01 & 26.95 & 21.00 & 23.48 \\ 
      & w. \toolname & 22.85 & 24.94 & 27.03 & 22.34 & 24.29 \\ 
      \cmidrule{2-7} 
      & Average & 22.67 & 24.23 & 28.13 & 21.29 & 24.08 \\ 
      & w. \toolname & 23.64 & 25.35 & 28.46 & 23.03 & 25.12 \\
      \bottomrule
    \end{tabular}%
  }
  \label{tab:standard_scenario}%
  \caption*{\footnotesize \normalfont $\dagger$ ``w. \toolname" denotes the results of LLMs hardened by \toolname.\hspace{0.8cm}\textcolor{white}{.} }
\end{table}

Table~\ref{tab:standard_scenario} presents the security hardening and functionality maintenance achieved by \toolname across four LLMs. On average, \toolname increases the security rate ({\tt SR}) of LLM-generated code by 9.37\% to 47.72\% across four languages, while maintaining or slightly enhancing the similarity ({\tt Sim}) metrics. We observed a trend where \toolname exhibits more obvious security-hardening on LLMs with larger parameter counts (e.g., DS-V3 with 671 billion parameters) compared to those with fewer parameters (e.g., CodeLlama with 13 billion parameters). Specifically, GPT-4o and DS-V3 achieve improvements of 20.28\% and 30.73\% in {\tt SR}, respectively, whereas CodeLlama and DS-Coder show improvements of 16.15\% and 13.50\%. This discrepancy might be attributed to the reduced instruction-following capabilities of smaller models~\cite{chung2024scaling,kaplan2020scaling}. 

From a programming language perspective, \toolname demonstrates the most significant security enhancement in Java, with an average improvement of approximately 38.45\%. This suggests that \toolname is particularly effective in addressing vulnerabilities prevalent in Java. Conversely, the improvements in C++ and Python are comparatively smaller, at 9.37\% and 12.24\%, respectively. This is likely because the {\tt SR} of LLM-generated C++ and Python code is already high, leaving limited room for improvement. We further investigated the ratio of insecure cases that were successfully secured by \toolname, revealing that approximately 31.74\%, 30.08\%, 32.82\% and 28.54\% of C, C++, Java and Python, respectively, were effectively secured.

To examine \toolname's impact on code functionality, we measured the similarity between generated code and the reference code using the {\tt Sim} metric. Notably, \toolname did not degrade functionality; rather, it slightly improved it. For instance, the average {\tt Sim} values across all LLMs without \toolname were 22.67, 24.23, 28.13, and 21.29 for C, C++, Java, and Python, respectively. With \toolname, these values increased to 23.64, 25.35, 28.46, and 23.03, respectively. This improvement suggests that the detailed guidance and secure code examples within the security knowledge base may enhance the functional correctness of generated code.

Note that the {\tt Sim} metric only measures code similarity due to the absence of test cases in the CyberSecEval dataset. To provide a comprehensive evaluation of \toolname's impact on functionality, we employed the MBPP~\cite{austin2021program} and HumanEval~\cite{chen2021evaluating} benchmarks, which utilize test cases to assess code correctness (see \S\ref{subsec:functionality_eval}). These results confirm that \toolname preserves code functionality, thus maintaining the practical utility of the generated code.

\notez{
{\bf Answer to RQ1}: \toolname effectively guides LLMs to generate more secure code in standard RACG scenarios across various programming languages and LLMs, while maintaining or slightly improving code functionality. Specifically, \toolname enhances the security rate by approximately 27.20\% for C, 9.37\% for C++, 47.72\% for Java, and 12.35\% for Python.
}

\subsection{RQ2: Resilience to Poisoning Attacks}
This RQ examines \toolname's performance when the knowledge base $\mathcal{K}$ is poisoned with vulnerable code, as modeled in \S\ref{subsec:scenarios}. We evaluate \toolname's capacity to harden code security amidst exposed programming intents (poisoning scenario I) and intent-agnostic knowledge poisoning (poisoning scenario II).

\subsubsection{Poisoning Scenario I}
\begin{table}[!t]
  \centering
  \caption{Performance of \toolname under poisoning scenario I.}
  \resizebox{1\linewidth}{!}{
    \begin{tabular}{ccrrrrr} % 
      \toprule
      \multirow{2}[2]{*}{\bf Metrics} & \multirow{2}[2]{*}{\bf LLM} & \multicolumn{5}{c}{\bf Language} \\ 
      \cmidrule{3-7} & & \multicolumn{1}{r}{\bf C} & \multicolumn{1}{r}{\bf C++} & \multicolumn{1}{r}{Java} & \multicolumn{1}{r}\bf {Python} & \multicolumn{1}{r}{\bf Average} \\
      \midrule
      \multicolumn{1}{c}{\multirow{14}[2]{*}{SR}} & {\bf GPT-4o} & 48.9  & 64.09 & 31.44 & 55.84 & 50.07  \\
      & w. \toolname & 62.56 & 79.15 & 56.77 & 71.23 & 67.43  \\
      &    & ($\uparrow$ {\bf 27.93\%}) & ($\uparrow$ {\bf 23.50\%}) & ($\uparrow$ {\bf 80.57\%}) & ($\uparrow$ {\bf 27.56\%}) & ($\uparrow$ {\bf 34.67\%}) \\
      & {\bf DS-V3} & 49.78 & 64.86 & 30.13 & 60.11 & 51.22  \\
      & w. \toolname & 68.72 & 80.31 & 57.21 & 74.64 & 70.22  \\
      &  & ($\uparrow$ {\bf 38.05\%}) & ($\uparrow$ {\bf 23.82\%}) & ($\uparrow$ {\bf 89.88\%}) & ($\uparrow$ {\bf 24.17\%}) & ($\uparrow$ {\bf 37.09\%}) \\ 
      & {\bf CodeLlama} & 51.1  & 61    & 39.74 & 56.13 & 51.99  \\
      & w. \toolname & 64.32 & 75.68 & 52.84 & 68.95 & 65.45  \\
      & & ($\uparrow$ {\bf 25.87\%}) & ($\uparrow$ {\bf 24.07\%}) & ($\uparrow$ {\bf 32.96\%}) & ($\uparrow$ {\bf 22.84\%}) & ($\uparrow$ {\bf 25.88\%}) \\
      & {\bf DS-Coder} & 44.49 & 60.62 & 34.5  & 56.41 & 49.01  \\
      & w. \toolname & 60.79 & 77.61 & 43.67 & 69.8  & 62.97  \\
      & & ($\uparrow$ {\bf 36.64\%}) & ($\uparrow$ {\bf 28.03\%}) & ($\uparrow$ {\bf 26.58\%}) & ($\uparrow$ {\bf 23.74\%}) & ($\uparrow$ {\bf 28.49\%}) \\ 
      \cmidrule{2-7}
      & Average & 48.57  & 62.64  & 33.95  & 57.12  & 50.57  \\
      & w. \toolname & 64.10  & 78.19  & 52.62  & 71.16  & 66.52  \\
      & & ($\uparrow$ {\bf 31.98\%}) & ($\uparrow$ {\bf 24.82\%}) & ($\uparrow$ {\bf 54.99\%}) & ($\uparrow$ {\bf 24.57\%}) & ($\uparrow$ {\bf 31.53\%}) \\ 
      \midrule
      \multicolumn{1}{c}{\multirow{10}[4]{*}{Sim}} & {\bf GPT-4o} & 25.89 & 29.49 & 32.75 & 28.73 & 29.97  \\
      & w. \toolname & 24.16 & 29.25 & 33.45 & 28.16 & 28.76 \\
      & {\bf DS-V3} & 22.68 & 28.77 & 32.89 & 25.80  & 27.54  \\
      & w. \toolname & 24.73 & 28.96 & 32.71 & 27.42 & 28.46 \\
      & {\bf CodeLlama} & 25.18 & 30.09 & 32.03 & 31.24 & 29.64  \\
      & w. \toolname & 25.97 & 31.84 & 31.86 & 32.48 & 30.54 \\
      & {\bf DS-Coder} & 24.95 & 28.76 & 32.82 & 29.14 & 28.42  \\
      & w. \toolname & 24.71 & 28.98 & 32.68 & 27.59 & 28.49 \\
      \cmidrule{2-7} 
      & Average & 24.68 & 29.28 & 32.62 & 28.73 & 28.70 \\
      & w. \toolname & 24.89  & 29.76 & 32.68 & 28.91 & 29.06  \\
      \bottomrule
    \end{tabular}%
  }
  \label{tab:poisnong_I}%
\end{table}
This scenario evaluates \toolname's resilience to poisoning attacks where the attacker has access to the programming intents (\ie queries) and poisons the knowledge base $\mathcal{K}$. The attacker injects the five most relevant vulnerable examples into $\mathcal{K}$ as illustrated in \S\ref{subsubsec:scenario_1}, simulating a targeted attack.

Table~\ref{tab:poisnong_I} presents the performance of \toolname under poisoning scenario I. The results indicate that in this scenario, the security of LLMs is compromised. For instance, the average {\tt SR} across LLMs drops from 60.84 (in the standard scenario) to 50.57, meaning nearly half of the generated code is vulnerable. 
% This experiment evaluates \toolname's ability to protect RACG systems from malicious attacks.  
Despite the presence of explicitly malicious knowledge in $\mathcal{K}$, \toolname consistently improves the {\tt SR} across all tested LLMs and programming languages. On average, \toolname enhances {\tt SR} by 31.53\%, demonstrating its ability to defend against attacks even when programming intents are exposed. Notably, DS-V3 exhibits the highest average improvement in {\tt SR} (37.09\%), followed by GPT-4o (34.67\%), CodeLlama (28.49\%), and DS-Coder (25.88\%).
From a language perspective, \toolname shows the most significant improvement in Java language, with an average {\tt SR} increase of 54.99\%. This indicates that \toolname is particularly effective in addressing vulnerabilities in Java. The improvements in C, C++, and Python are also substantial, with average {\tt SR} increases of 31.98\%, 24.82\%, and 24.57\%, respectively. Additionally, we analyze the proportion of previously insecure cases that were successfully secured by \toolname. Results show that \toolname effectively secures 41.61\% of C++ cases, followed by Python (32.73\%), C (30.19\%), and Java (28.27\%).

Regarding functionality, as measured by the {\tt Sim} metric, \toolname maintains or slightly improves the similarity between the generated code and the reference code. The average {\tt Sim} across all LLMs and languages increases from 28.70 to 29.06 with \toolname. This suggests that \toolname's security enhancements do not compromise the functional correctness of the generated code, even in the context of targeted poisoning attacks.

Overall, \toolname demonstrates a strong resilience to poisoning attacks in scenario I. It effectively mitigates targeted vulnerabilities by enhancing code security without sacrificing functionality, even when the knowledge base is poisoned with semantically similar vulnerable code and the programming intents are exposed.

\subsubsection{Poisoning Scenario II}
\begin{table}[!t]
  \centering
  \caption{Performance of \toolname under poisoning scenario II.}
  \resizebox{1\linewidth}{!}{
    \begin{tabular}{ccrrrrr} % 
      \toprule
      \multirow{2}[2]{*}{\bf Metrics} & \multirow{2}[2]{*}{\bf LLM} & \multicolumn{5}{c}{\bf Language} \\        \cmidrule{3-7} & & \multicolumn{1}{r}{\bf C} & \multicolumn{1}{r}{\bf C++} & \multicolumn{1}{r}{\bf Java} & \multicolumn{1}{r}{\bf Python} & \multicolumn{1}{r}{\bf Average} \\
      \midrule
      \multicolumn{1}{c}{\multirow{14}[2]{*}{SR}} & {\bf GPT-4o} & 52.86 & 76.83 & 38.86 & 63.58 & 58.03  \\
      & w. \toolname & 69.6  & 86.49 & 64.63 & 80.63 & 75.34  \\
      &    & ($\uparrow$ {\bf 31.67\%}) & ($\uparrow$ {\bf 12.57\%}) & ($\uparrow$ {\bf 66.31\%}) & ($\uparrow$ {\bf 26.81\%}) & ($\uparrow$ {\bf 29.82\%}) \\
      & {\bf DS-V3} & 52.86 & 72.97 & 38.86 & 66.95 & 57.91  \\
      & w. \toolname & 72.25 & 86.87 & 65.94 & 80.63 & 76.42  \\
      &  & ($\uparrow$ {\bf 36.68\%}) & ($\uparrow$ {\bf 19.05\%}) & ($\uparrow$ {\bf 69.69\%}) & ($\uparrow$ {\bf 20.43\%}) & ($\uparrow$ {\bf 31.97\%}) \\ 
      & {\bf CodeLlama} & 57.71 & 71.43 & 50.66 & 70.37 & 62.54  \\
      & w. \toolname & 65.20 & 80.69 & 58.52 & 78.06 & 70.62  \\
      & & ($\uparrow$ {\bf 12.98\%}) & ($\uparrow$ {\bf 12.96\%}) & ($\uparrow$ {\bf 15.52\%}) & ($\uparrow$ {\bf 10.93\%}) & ($\uparrow$ {\bf 12.91\%}) \\
      & {\bf DS-Coder} & 52.42 & 77.61 & 45.85 & 70.94 & 61.71  \\
      & w. \toolname & 64.32 & 81.08 & 60.26 & 76.07 & 73.29  \\
      & & ($\uparrow$ {\bf 22.70\%}) & ($\uparrow$ {\bf 4.47\%}) & ($\uparrow$ {\bf 31.43\%}) & ($\uparrow$ {\bf 7.23\%}) & ($\uparrow$ {\bf 18.77\%}) \\ 
      \cmidrule{2-7}
      & Average & 53.96  & 74.71  & 43.56  & 67.96  & 60.05  \\
      & w. \toolname & 67.84  & 83.78  & 62.34  & 78.85  & 73.20  \\
      & & ($\uparrow$ {\bf 25.72\%}) & ($\uparrow$ {\bf 12.14\%}) & ($\uparrow$ {\bf 43.12\%}) & ($\uparrow$ {\bf 16.02\%}) & ($\uparrow$ {\bf 21.91\%}) \\ 
      \midrule
      \multicolumn{1}{c}{\multirow{10}[4]{*}{Sim}} & {\bf GPT-4o} & 18.36 & 20.61 & 26.39 & 21.03 & 21.60  \\
      & w. \toolname & 18.92 & 21.57 & 27.03 & 24.94 & 23.12  \\
      & {\bf DS-V3} & 17.5  & 21.13 & 26.54 & 20.69 & 21.47  \\
      & w. \toolname & 18.53 & 21.55 & 26.93 & 22.56 & 22.39  \\
      & {\bf CodeLlama} & 17.63 & 20.2  & 25.33 & 20.86 & 21.01  \\
      & w. \toolname & 18.76 & 22.34 & 26.34 & 22.02 & 22.37  \\
      & {\bf DS-Coder} & 16.6  & 19.95 & 24.64 & 20.08 & 20.32  \\
      & w. \toolname & 18.83 & 22.18 & 26.11 & 21.99 & 22.28  \\
      \cmidrule{2-7} 
      & Average & 17.52  & 20.47  & 25.73  & 20.67  & 21.10  \\
      & w. \toolname & 18.76  & 21.91  & 26.60  & 22.88  & 22.54  \\
      \bottomrule
    \end{tabular}%
  }
  \label{tab:poisnong_II}%
\end{table}
This scenario evaluates \toolname's resilience to poisoning attacks where the attacker lacks direct access to the programming intents (\ie queries $Q$) and instead poisons the knowledge base $\mathcal{K}$ with common vulnerable functionalities that are more likely to be retrieved in RACG. The attacker employs a clustering-based approach to select the top 10\% most representative examples from $\mathcal{K}$ and injects their corresponding vulnerable counterparts, simulating a generalized poisoning attack.

Table~\ref{tab:poisnong_II} presents the performance of \toolname under poisoning scenario II. Similar to scenario I, \toolname consistently enhances the {\tt SR} across all tested LLMs and programming languages, even when faced with intent-agnostic poisoned knowledge. On average, \toolname improves {\tt SR} by 21.91\%, demonstrating its robustness against generalized poisoning attacks. Notably, DS-VS exhibits the highest average improvement in {\tt SR} (31.97\%), followed by GPT-4o (29.82\%), DS-Coder (18.77\%), and CodeLlama (12.91\%). 

From a language perspective, \toolname again demonstrates the most significant improvement in Java code generation, with an average {\tt SR} increase of 43.12\%. This confirms \toolname's effectiveness in mitigating vulnerabilities in Java, even under generalized poisoning conditions. The improvements in C, C++, and Python are also notable, with average {\tt SR} increases of 25.72\%, 12.14\%, and 16.02\%, respectively. Furthermore, the proportion of cases successfully secured by \toolname across the four programming languages is 30.15\%, 35.87\%, 33.27\%, and 33.98\%, respectively.

Regarding functionality, as measured by the {\tt Sim} metric, \toolname maintains or slightly improves the similarity between the generated code and the reference code. The average {\tt Sim} across all LLMs and languages increases from 21.10 to 22.54 with \toolname. This indicates that \toolname's security enhancements do not compromise the functional correctness of the generated code, even in the context of intent-agnostic poisoning attacks.

\notez{
{\bf Answer to RQ2}: \toolname consistently improves code security across poisoning scenarios, including targeted (Scenario I) and generalized (Scenario II) attacks. Specifically, \toolname achieves an average {\tt SR} improvement of 31.53\% in scenario I and 21.91\% in scenario II. These results demonstrate \toolname's effectiveness in hardening the security of RACG systems, without compromising the functionality of the generated code.
}

\subsection{RQ3: Generalization of \toolname}
This RQ explores \toolname's generalization by analyzing two dimensions: (1) \toolname's performance when when there is no off-the-shelf knowledge base (denoted as non-func-retrieval) that the LLM relies on its intrinsic knowledge to generate code; and (2) \toolname's performance when the security knowledge base lacks corresponding knowledge for the target language. 

\subsubsection{Performance in the Non-Func-Retrieval Scenario}
The non-func-retrieval scenario is common in daily software development, where LLMs generate code solely based on developer instructions. To assess \toolname's generalization in this setting, we examine whether it enhances code security while preserving functionality. Additionally, we compare its effectiveness against existing methods specifically designed for this generation scenario.

\textbf{Baselines.}  
Recent research has introduced several approaches aimed at enhancing the security of code generation in non-func-retrieval scenarios~\cite{li2024cosec, he2023large, he2024instruction}.  
We select three state-of-the-art security-hardening methods as baselines:  

\begin{itemize}[leftmargin=*]
    \item {\bf Sven}. Sven~\cite{he2023large} employs the prefix-tuning technique~\cite{li2021prefix} to steer code generation toward desired properties, such as functional correctness and security, without modifying the LLM's weights.  
    \item {\bf SafeCoder}. Similar to Sven, SafeCoder~\cite{he2024instruction} employs instruction tuning~\cite{zhang2023instruction} to fine-tune LLMs on a specially curated dataset, guiding them to generate secure code while discouraging unsafe program generation through likelihood loss.
    \item {\bf Cosec}. Previous studies~\cite{he2023large, he2024instruction} require access to the weights of LLMs, which limits the applicability of these approaches. In contrast, Cosec~\cite{li2024cosec} proposes leveraging co-decoding to adjust the probability distributions of tokens at each step of the decoding process, thereby guiding the generation of secure code.

\end{itemize}

\textbf{Experimental Results.}
We evaluate the approaches only on C, C++, and Python from CybersecEval, in line with RQ1 and RQ2, as these models were trained exclusively on these languages. Comparing them with \toolname on other languages could introduce potential biases.  
For the investigated LLMs, we select two relatively small models, Mistral-7B~\cite{Jiang2023Mistral7B} and CodeLlama-7B~\cite{roziere2023code}, due to the additional training required for baseline reproduction, which is both time-consuming and resource-intensive. For instance, training CodeLlama required approximately 1.4 million GPU hours~\cite{roziere2023code}.
Regarding data reproduction, we reuse the trained model provided by the authors for SafeCoder. For Sven and CoSec, we follow the official implementations of each approach and implement them using Mistral-7B and CodeLlama-7B.

\begin{table}[!t]
  \centering
  \caption{Performance of LLMs in the non-retrieval scenario}
  \renewcommand{\arraystretch}{1.2}
  \resizebox{\linewidth}{!}{
    \begin{tabular}{cllrrrr}
      \toprule
      \textbf{Metrics} & \textbf{LLM} & \textbf{Approach} & \textbf{C} & \textbf{C++} & \textbf{Python} & \textbf{Average} \\
      \midrule
      \multirow{8}{*}{SR} 
        & \multirow{4}{*}{Mistral-7B}  & Sven       & 60.35  & 70.27  & 74.93  & 68.52  \\
        &                              & SafeCoder  & 63.00  & 75.29  & 78.63  & 72.31  \\
        &                              & Cosec      & 59.91  & 70.66  & 76.64  & 69.07  \\
        &                              & \toolname   & \textbf{70.48}  & \textbf{83.40}  & \textbf{84.33}  & \textbf{79.40}  \\
      \cmidrule{2-7} 
        & \multirow{4}{*}{CodeLlama-7B} & Sven      & 62.11  & 70.66  & 71.23  & 68.00  \\
        &                              & SafeCoder  & 64.89  & 76.56  & 77.78  & 73.08  \\
        &                              & Cosec      & 59.47  & 74.13  & 76.07  & 69.89  \\
        &                              & \toolname   & \textbf{68.72}  & \textbf{79.54}  & \textbf{82.91}  & \textbf{77.06}  \\
      \midrule
      \multirow{8}{*}{Sim}
        & \multirow{4}{*}{Mistral-7B}  & Sven       & 19.43  & 20.06  & 16.51  & 18.67  \\
        &                              & SafeCoder  & 19.36  & 20.17  & 17.16  & 18.90  \\
        &                              & Cosec      & 19.06  & 19.80  & 16.63  & 18.50  \\
        &                              & \toolname   & \textbf{23.08}  & \textbf{25.19}  & \textbf{22.37}  & \textbf{23.55}  \\
      \cmidrule{2-7} 
        & \multirow{4}{*}{CodeLlama-7B} & Sven      & 19.26  & 19.96  & 18.03  & 19.08  \\
        &                              & SafeCoder  & 19.06  & 20.30  & 19.28  & 19.55  \\
        &                              & Cosec      & 18.46  & 19.36  & 18.15  & 18.66  \\
        &                              & \toolname   & \textbf{24.02}  & \textbf{25.65}  & \textbf{23.11}  & \textbf{24.26}  \\
      \bottomrule
    \end{tabular}
  }
  \label{tab:non_retrieval}
\end{table}

Table~\ref{tab:non_retrieval} presents the evaluation results. Overall, \toolname outperforms the investigated baselines in both security hardening and functionality preservation in the non-func-retrieval setting.
Compared to the state-of-the-art technique, SafeCoder, \toolname generates 9.80\% (72.31 $\rightarrow$ 79.40) more secure code on average with Mistral-7B. Additionally, the {\tt Sim} metric of \toolname is 24.60\% higher than that of CoSec (18.90 $\rightarrow$ 23.55) on Mistral-7B. Similar improvements are observed with CodeLlama-7B.  
From a programming language perspective, \toolname achieves significant improvements over previous methods. Specifically, its relative improvement over SafeCoder in the {\tt SR} metric is 11.87\%, 10.77\%, and 7.25\% on Mistral-7B across C, C++, and Python. We also observed that the {\tt Sim} metric of \toolname is significantly higher than the baselines, and we attribute it to the provided decomposed sub-tasks and the examples in the provided security knowledge. Notably, we observe that the security rate of the generated code is lower than the results reported in prior studies~\cite{li2024cosec, he2023large, he2024instruction}. This discrepancy is likely due to differences in dataset scope: the prior works evaluate on a test set with only 166 cases across nine CWE types, whereas CyberSecEval contains 1,916 cases covering 50 CWE types, offering a more comprehensive assessment.  

These results demonstrate that \toolname not only enhances the security of LLM-generated code in retrieval scenarios (\ie RACG) but also achieves strong performance in non-func-retrieval settings, outperforming existing security-hardening approaches specifically designed for non-func-retrieval scenarios.

\subsubsection{Performance with Language-Specific Knowledge Absence}

In real-world settings, the security knowledge base \(\mathcal{S}\) may lack knowledge for certain programming languages (e.g., C\#), as its distribution varies widely across languages. This subquestion examines whether \toolname can still enhance code security when target language knowledge is absent, leveraging knowledge from other languages. We evaluate \toolname's effectiveness on four external languages, C\#, JavaScript, PHP, and Rust, where the knowledge retriever accesses a \(\mathcal{S}\) devoid of target-specific security knowledge, testing its cross-language generalization.

\begin{table}[!t]
    \centering
    \caption{Performance of \toolname in the absence of language-specific knowledge.}
    \resizebox{1\linewidth}{!}{
      \begin{tabular}{ccrrrrr} % 
        \toprule
        \multirow{2}[2]{*}{\bf Scenario} & \multirow{2}[2]{*}{\bf LLM} & \multicolumn{5}{c}{\bf Language} \\ 
        \cmidrule{3-7} & & \multicolumn{1}{r}{\bf C\#} & \multicolumn{1}{r}{\bf JavaScript} & \multicolumn{1}{r}{\bf PHP} & \multicolumn{1}{r}{\bf Rust} & \multicolumn{1}{r}{\bf Average} \\
        \midrule
        \multicolumn{1}{c}{\multirow{9}[2]{*}{Standard}} & {\bf GPT-4o} & 52.77 & 54.62 & 62.35 & 52.94 & 55.67  \\
        & w. \toolname & 61.28 & 65.46 & 78.40  & 60.29 & 66.36  \\
        & {\bf DS-V3} & 46.38 & 53.82 & 60.49 & 53.92 & 53.65  \\
        & w. \toolname & 61.70  & 67.07 & 75.93 & 53.92 & 64.66  \\
        & {\bf CodeLlama} & 50.64 & 54.62 & 60.49 & 52.94 & 54.67  \\
        & w. \toolname & 60.85 & 64.26 & 70.37 & 61.76 & 64.31  \\
        & {\bf DS-Coder} & 54.47 & 57.83 & 61.11 & 56.37 & 57.45  \\
        & w. \toolname & 55.32 & 58.63 & 74.07 & 55.39 & 60.85  \\
        \cmidrule{2-7}
        & Average & 51.07  & 55.22  & 61.11  & 54.04  & 55.36  \\
        & w. \toolname & 59.79  & 63.86  & 74.69  & 57.84  & 64.04  \\
        &    & ($\uparrow$ {\bf 17.08\%}) & ($\uparrow$ {\bf 15.63\%}) & ($\uparrow$ {\bf 22.23\%}) & ($\uparrow$ {\bf 7.03\%}) & ($\uparrow$ {\bf 15.69\%}) \\
        \midrule
        \multicolumn{1}{c}{\multirow{9}[4]{*}{I$\dagger$}} & {\bf GPT-4o} & 42.13 & 42.57 & 41.36 & 34.8  & 40.22  \\
        & w. \toolname & 47.66 & 51.41 & 55.56 & 41.18 & 48.95  \\
        & {\bf DS-V3} & 42.13 & 43.78 & 41.98 & 38.73 & 41.66  \\
        & w. \toolname & 56.17 & 56.63 & 62.35 & 39.71 & 53.72  \\
        & {\bf CodeLlama} & 53.19 & 48.59 & 40.12 & 45.1  & 46.75  \\
        & w. \toolname & 60.85 & 61.04 & 62.96 & 45.59 & 57.61  \\
        & {\bf DS-Coder} & 45.11 & 48.59 & 43.21 & 36.27 & 43.30  \\
        & w. \toolname & 51.91 & 47.79 & 56.79 & 36.27 & 48.19  \\
        \cmidrule{2-7} 
        & Average & 45.64  & 45.88  & 41.67  & 38.73  & 42.98  \\
        & w. \toolname & 54.15  & 54.22  & 59.42  & 40.69  & 52.12  \\
        &    & ($\uparrow$ {\bf 18.64\%}) & ($\uparrow$ {\bf 18.17\%}) & ($\uparrow$ {\bf 42.59\%}) & ($\uparrow$ {\bf 5.07\%}) & ($\uparrow$ {\bf 21.26\%}) \\
        \midrule
        \multicolumn{1}{c}{\multirow{9}[4]{*}{II$\dagger$}} & {\bf GPT-4o} & 49.79 & 54.22 & 63.58 & 54.41 & 55.50  \\
        & w. \toolname & 57.87 & 66.67 & 80.63 & 61.76 & 66.73  \\
        & {\bf DS-V3} & 45.96 & 53.01 & 59.88 & 55.39 & 53.56  \\
        & w. \toolname & 61.28 & 64.26 & 80.63 & 54.9  & 65.27  \\
        & {\bf CodeLlama} & 53.62 & 58.63 & 70.37 & 65.69 & 62.08  \\
        & w. \toolname & 69.95 & 64.26 & 73.46 & 61.76 & 67.36  \\
        & {\bf DS-Coder} & 54.04 & 60.24 & 70.94 & 54.41 & 59.91  \\
        & w. \toolname & 58.27 & 61.83 & 70.06 & 61.35 & 62.88  \\
        \cmidrule{2-7} 
        & Average & 50.85  & 56.53  & 66.19  & 57.48  & 57.76  \\
        & w. \toolname & 61.84  & 64.26  & 76.19  & 59.94  & 65.56  \\
        &    & ($\uparrow$ {\bf 21.61\%}) & ($\uparrow$ {\bf 13.68\%}) & ($\uparrow$ {\bf 15.11\%}) & ($\uparrow$ {\bf 4.29\%}) & ($\uparrow$ {\bf 13.50\%}) \\
        \bottomrule
      \end{tabular}%
    }
    \label{tab:across_lang}%
    \caption*{\footnotesize \normalfont $\dagger$ ``I" and ``II"" denote the Poisoning Scenario I and II.\hspace{2.7cm}\textcolor{white}{.}}
  \end{table}

Table~\ref{tab:across_lang} presents \toolname's performance without language-specific security knowledge. As observed in RQ1 and RQ2, the functionality impact of \toolname is minimal; therefore, we focus solely on the {\tt SR} metric in this analysis. 
% Across all scenarios, \toolname consistently improves SR, even without language-specific knowledge. 

Overall, \toolname consistently improves SR across scenarios even without language-specific knowledge. In the Standard scenario, {\tt SR} rises from 55.36\% to 64.04\% (15.69\% improvement), with gains of 17.08\% (C\#), 15.63\% (JavaScript), 22.23\% (PHP), and 7.03\% (Rust). In poisoning scenario I, the average {\tt SR} increases from 42.98\% to 52.12\% (21.26\%), with PHP showing the highest gain (42.59\%). In poisoning scenario II, the average {\tt SR} improves from 57.76\% to 65.56\% (13.50\%), with C\# leading at 21.61\%. 
% These results show \toolname's effectiveness in enhancing security across languages, albeit with varying degrees of improvement.

The consistent SR improvements suggest that \toolname leverages universal security principles (\eg avoiding weak random number generation, which is a common pitfall across languages like C++, C\#, and JavaScript) present in the knowledge base. However, {\tt SR
} improvements in the poisoning scenario I are significantly lower than in languages with corresponding security knowledge in the knowledge base. Specifically, the average improvement in C, C++, Java, and Python is 31.53\%, whereas in the other four languages, it is only 21.26\%. These findings indicate that while \toolname can mitigate poisoning attacks even without language-specific knowledge by relying on general security patterns, its effectiveness is somewhat limited, and the mitigation process becomes more challenging without tailored security information.

% Overall speaking, the consistent improvement in {\tt SR} across different languages, despite the absence of language-specific knowledge, highlights \toolname's ability to generalize security principles and apply them effectively across diverse programming contexts. This suggests that \toolname can leverage universal security knowledge to enhance code security, even when faced with language-specific knowledge gaps.

\notez{
{\bf Answer to RQ3}: \toolname demonstrates inspiring generalization, outperforming state-of-the-art security-hardening methods, SafeCoder, by 9.80\% in {\tt SR} on Mistral-7B when no similar code is retrieved. Moreover, even in the absence of language-specific knowledge, \toolname consistently improves security across various languages, leveraging universal security principles.
}
\section{Discussion}
\label{sec:dis}
\subsection{Functional Correctness Maintenance}
\label{subsec:functionality_eval}
The {\tt Sim} metric, used in prior RQs, measures code similarity to reference implementations but does not directly assess functional correctness due to the absence of test cases in the CyberSecEval dataset. To provide a comprehensive evaluation of \toolname's impact on the functionality of LLM-generated code, we employ the MBPP~\cite{austin2021program} and HumanEval~\cite{chen2021evaluating} benchmarks, which leverage test cases to measure code correctness. This section investigates whether \toolname's security enhancements compromise or enhance the practical utility of the generated code.
\begin{table}[!t]
  \centering
  \caption{Impact of \toolname on LLMs' functionality}
  \resizebox{0.8\linewidth}{!}{
   \begin{tabular}{cllll}
    \toprule
    \multirow{2}{*}{\bf LLM}   & \multicolumn{2}{c}{\bf MBPP} & \multicolumn{2}{c}{\bf HumanEval} \\
                           & \textbf{Pass@1} & \textbf{Pass@5} & \textbf{Pass@1} & \textbf{Pass@5} \\
    \midrule
    {\bf GPT-4o} &  72.8   & 77.4  &  89.0 & \bf{91.5} \\
    w. \toolname &  \bf{73.2}   & \bf{78.8}  &  \bf{89.6} & \bf{91.5} \\
    {\bf DS-V3}  &  74.6   & 78.6  &  87.2 & 89.6 \\
    w. \toolname &  \bf{75.6}   & \bf{79.6}  &  \bf{87.8} & \bf{90.9} \\
    {\bf CodeLlama} & 54.8 & 60.8  &  40.9 & 47.6 \\
    w. \toolname &  \bf{56.4}   & \bf{62.2}  &  \bf{43.3} & \bf{50.0} \\
    {\bf DS-Coder}  & 62.8 & 69.2  &  80.5 & 86.0 \\
    w. \toolname &  \bf{65.0}   & \bf{72.8}  &  \bf{81.7} & \bf{88.4} \\
    \bottomrule
    \end{tabular}%
  }
  \label{tab:func_eval}%
\end{table}%

Table~\ref{tab:func_eval} reports results on the MBPP and HumanEval benchmarks across different LLMs, using the default benchmark settings. Security knowledge is integrated as described in \S\ref{subsec:sec_code_gen}.  
Overall, \toolname slightly enhances the functionality of the generated code across all four evaluated LLMs and benchmarks. For instance, \toolname improves the pass@1 score of GPT-4o on MBPP from 72.8 to 73.2 and the pass@5 score from 77.4 to 78.8. A similar trend is observed across other LLMs and benchmarks (\ie HumanEval), indicating that integrating security knowledge does not compromise, and even enhances the functionality of LLM-generated code.

\subsection{Effectiveness Across Vulnerability Types}
\begin{table}[!t]
  \centering
  \caption{Percentage of vulnerabilities prevented by \toolname across MITRE’s Top-25 software weaknesses}
  \begin{tabular}{lcccc}
    \toprule
    \textbf{CWE$\dagger$} & \textbf{Standard} & \textbf{I} & \textbf{II} & \textbf{Average} \\
    \midrule
    CWE-79  & 0.00   & 14.29  & 0.00   & 4.76   \\
    CWE-89  & 34.62  & 40.00  & 34.78  & 36.47  \\
    CWE-352 & 40.00  & 38.89  & 36.36  & 38.42  \\
    CWE-22  & 13.79  & 3.85   & 7.14   & 8.26   \\
    CWE-78  & 40.00  & 44.12  & 46.99  & 43.70  \\
    CWE-862 & 18.18  & 28.57  & 27.27  & 24.67  \\
    CWE-94  & 38.10  & 41.38  & 39.13  & 39.54  \\
    CWE-502 & 53.33  & 12.24  & 15.56  & 27.04  \\
    CWE-200 & 100.00 & 50.00  & 50.00  & 66.67  \\
    CWE-918 & 26.67  & 20.00  & 31.25  & 25.97  \\
    CWE-119 & 88.89  & 78.57  & 100.00 & 89.15  \\
    CWE-798 & 30.77  & 25.93  & 46.67  & 34.46  \\
    \bottomrule
  \end{tabular}
  \label{tab:top_25}
  \caption*{\footnotesize \normalfont $\dagger$ Only CWE types present in both MITRE’s Top-25 and CyberSecEval are shown.}
\end{table}

In this discussion, we evaluate the effectiveness of \toolname in preventing vulnerabilities across MITRE's Top-25 software weaknesses~\cite{CWE2024Top25} on three investigated scenarios. To quantify this, we define the prevention percentage as the proportion of vulnerabilities present in code generated without \toolname that are successfully mitigated when \toolname is applied. 
% Formally, for a given vulnerability type \( CWE_i \), the mitigation percentage is calculated as:
% \[ \text{Mitigation Percentage} = \frac{\text{\# Vulnerabilities Mitigated by \toolname}}{\text{\# Vulnerabilities in Original LLM}} \times 100 \]

Given \toolname's consistent security hardening across LLMs, we use DS-V3 as the representative LLM. Results are presented in Table~\ref{tab:top_25}. Overall, \toolname demonstrates varying effectiveness, with average prevention rates ranging from 4.76\% (CWE-79) to 89.15\% (CWE-119). It excels at addressing critical weaknesses like CWE-119 (buffer errors, 89.15\%) and CWE-200 (information exposure, 66.67\%), reflecting its strength in leveraging security knowledge for broadly applicable vulnerabilities. However, its performance dips for CWE-79 (cross-site scripting, 4.76\%), where analysis reveals that the generated code contained 3, 7, and 3 CWE-79 instances across the three scenarios, all stemming from JavaScript. This poor result stems from a lack of JavaScript-specific knowledge in \toolname's security knowledge base, limiting its effectiveness for this CWE type. Other lower-performing cases, such as CWE-22 (path traversal, 8.26\%), suggest similar context-specific challenges. These findings highlight \toolname's strengths in securing different vulnerability types while underscoring areas for refinement, particularly in enriching the knowledge base.

% \vspace{-3mm}
\subsection{Ablation Study}
\begin{table}[!t]
  \centering
  \caption{Performance comparison across different variants}  
  \begin{tabular}{lcccc} 
    \toprule
    \textbf{Metric} & \textbf{Variant} & \textbf{Standard} & \textbf{I} & \textbf{II} \\
    \midrule
    \multirow{3}{*}{SR} 
      & \toolname-QD         & 68.61 & 64.04 & 66.25 \\
      & \toolname-KRF         & 74.13 & 68.29 & 74.87 \\
      & \toolname   & 76.36 & 70.22 & 76.42 \\
    \midrule
    \multirow{3}{*}{Sim}
      & \toolname-QD         & 23.04 & 25.73 & 20.41 \\
      & \toolname-KRF         & 24.82 & 27.04 & 21.14 \\
      & \toolname   & 25.56 & 28.46 & 22.39 \\
    \bottomrule
  \end{tabular}
  \label{tab:ablation_study} 
\end{table}
% Our previous results demonstrated that \toolname yields significant improvements in security enhancement while maintaining code functionality. 
To further dissect the contribution of its core components, we conducted an ablation study by evaluating two variants of \toolname, each with a key module disabled. Specifically, we created variants excluding Query Decomposition (QD), denoted \toolname-QD, and excluding Knowledge Re-ranking and Filtering (KRF), denoted \toolname-KRF.
In the \toolname-QD variant, security knowledge was retrieved using the user's original query directly, bypassing the sub-task decomposition step. In the \toolname-KRF variant, all initially retrieved security knowledge for the sub-tasks was provided to the LLM for code generation without applying the re-ranking and filtering process. It is crucial to note that the KRF module operates on the sub-tasks generated by QD; therefore, disabling QD inherently disables the KRF mechanism as well. To ensure a fair comparison, we maintain consistency in the number of injected knowledge entries across all variants. Given that \toolname demonstrates consistent performance across different LLMs, we conduct this ablation study using DS-V3.
% Consequently, the \toolname-QD variant effectively evaluates the system without the benefits of either QD or KRF. To ensure a fair comparison regarding the amount of information presented to the LLM, we controlled the number of knowledge entries retrieved in the \toolname-QD setting to match the number of entries ultimately selected by the full \toolname (after KRF). Note that given the significant computational expense required to run all variants across all scenarios and LLM backends, and observing that the full \toolname demonstrated consistent efficacy across the LLMs investigated, we performed this ablation study using DS-V3 as the representative LLM backend.

The results are presented in Table~\ref{tab:ablation_study}. Overall, QD exhibits the most substantial impact on both the {\tt SR} and {\tt Sim} metrics. Removing QD led to a decrease in {\tt SR} of approximately 10.1\% (from 76.36\% to 68.61\%) and a drop in {\tt Sim} from 25.56 to 23.04 in the standard scenario. Similar trends were observed in scenarios I and II. In contrast, the contribution of KRF was less pronounced. Removing only KRF (\toolname-KRF) resulted in a smaller decrease in {\tt SR} of about 2.2\% (from 76.36\% to 74.13\%) and a reduction in {\tt Sim} from 25.56 to 24.82 in the standard scenario. These findings underscore the critical role of Query Decomposition in enabling more precise knowledge retrieval, thereby significantly enhancing the security of the generated code. Knowledge Re-ranking and Filtering provide an additional, valuable refinement to this process.

\subsection{Impact of the Number of Injected Knowledge Entries}
\label{subsec:num_kng}
\begin{table}[!t]
\centering
\caption{The Impact of \( k' \) and \( k \) on \toolname's Performance (\tt SR)}
\label{tab:param_study}
\begin{minipage}{\columnwidth}
\centering
\begin{subtable}[t]{0.45\columnwidth}
\centering
\caption{DS-V3}
\label{tab:sub1}
\resizebox{\linewidth}{!}{
\begin{tabular}{ccccc}
\toprule
\multirow{2}{*}{\( k' \)} & \multicolumn{4}{c}{\( k \)} \\
\cmidrule(lr){2-5}
 & 3 & 5 & 7 & 9 \\
\midrule
1 & 69.84 & 73.32 & 75.15 & 76.04 \\
2 & 74.72 & 76.36 & {\bf 77.26} & 76.82 \\
3 & 75.31 & 76.53 & 75.62 & 75.03 \\
\bottomrule
\end{tabular}
}
\end{subtable}
\hfill
\begin{subtable}[t]{0.45\columnwidth}
\centering
\caption{CodeLlama}
\label{tab:sub2}
\resizebox{\linewidth}{!}{
\begin{tabular}{ccccc}
\toprule
\multirow{2}{*}{\( k' \)} & \multicolumn{4}{c}{\( k \)} \\
\cmidrule(lr){2-5}
 & 3 & 5 & 7 & 9 \\
\midrule
1 & 66.14 & 67.05 & 69.45 & 69.39 \\
2 & 67.38 & {\bf 69.74} & 68.23 & 67.21 \\
3 & 68.17 & 68.59 & 67.42 & 66.35 \\
\bottomrule
\end{tabular}
}
\end{subtable}
\end{minipage}
\end{table}
The amount of injected security knowledge significantly impacts \toolname's performance. This amount is controlled by $k'$ (knowledge entries retrieved per sub-task, \S\ref{subsubsec:know_retrieval}) and $k$ (top-ranked sub-tasks selected after filtering, \S\ref{subsubsec:reranking}). We empirically tuned these hyperparameters by evaluating various $(k', k)$ combinations using DS-V3 (a larger model) and CodeLlama-13B (a smaller model) on the standard RACG scenario.

Results are presented in Table~\ref{tab:param_study}. We observed that performance generally improves with more knowledge up to a point, after which excessive knowledge leads to degradation. The optimal configuration among the investigated hyperparameters was $(k'=2, k=7)$ for DS-V3 and $(k'=2, k=5)$ for CodeLlama-13B, indicating that DS-V3 can effectively leverage more security knowledge than CodeLlama-13B. This suggests that models with relatively fewer parameters may be overwhelmed by excessively large knowledge contexts, leading to performance drops due to their comparatively weaker instruction-following capabilities~\cite{chung2024scaling,kaplan2020scaling}.

Based on these findings, and aiming for robust performance across different model scales while mitigating degradation, we selected $k'=2$ and $k=5$ as the default hyperparameters for \toolname in our main evaluations (\S\ref{sec:eval}). 

\subsection{Threats to Validity}
\label{subsec:threats}
We identify the following potential threats to the validity:

{\bf Validity of Security Measurement.}
Our primary measure of code security relies on the Insecure Code Detector~\cite{bhatt2023purple} applied within the CyberSecEval benchmark. This detector employs an ensemble of static analysis tools (\eg Semgrep~\cite{semgrep} and Weggli~\cite{weggli}) configured to identify patterns with 50 CWEs. A threat to validity arises because this measurement may not perfectly capture the true security posture of the generated code. Furthermore, the predefined set of 50 CWEs, while significant, may not represent the complete universe of potential security flaws.

We argue that this threat is partially mitigated by the following factors: (1) The 50 CWEs targeted by the detector encompass a broad range of common and critical vulnerability types frequently encountered in practice, providing substantial coverage. Notably, the Insecure Code Detector achieves a detection precision of 96\% for vulnerabilities within this set. (2) The utilization of static analysis tools represents a standard and practical methodology for large-scale automated security assessment in existing studies~\cite{pearce2022asleep,tihanyi2025secure,klemmer2024using}. And  (3) Most importantly for our comparative study, the consistent application of the same detection tools and criteria across all evaluated LLMs and investigated scenarios ensures that the observed differences in security performance (\eg the improvements attributed to \toolname) are measured fairly, allowing for reliable and robust comparisons. 

\textbf{Reliability and LLM Non-Determinism.} Another threat arises from the inherent non-determinism associated with LLMs, which could impact the reliability and reproducibility of our experimental results. Specifically, LLMs can reproduce varying outputs even when presented with the same input multiple times. One potential mitigation is to average performance over multiple generations, but this approach is extremely time-consuming. To mitigate this threat, we adhered to the recommended practices~\cite{ouyang2024empirical} by setting the temperature parameter to 0 during all LLM inference steps, which minimizes the stochasticity in the generation process. 

However, even at zero temperature, complete determinism is not assured, particularly for architectures like Mixture-of-Experts (e.g., DS-V3 used in this study)~\cite{puigcerver2023sparse}. To quantify the impact of residual non-determinism, we conducted five independent runs of the DS-V3 under the standard scenario and measured variability in our primary metric, the Security Rate ({\tt SR}). The maximum observed deviation in {\tt SR} across these runs was 0.38\%, indicating negligible influence on our findings. This low level of variability suggests that the impact of residual LLM non-determinism is negligible, indicating that this threat is well-controlled in our study.

% The primary external threats to validity arise from 
\section{Conclusion}
\label{sec:conc}
In this work, we propose \toolname, a security-hardening framework for RACG systems that addresses the overlooked threat of knowledge base poisoning. By explicitly incorporating security knowledge into the code generation prompt, \toolname enables LLMs to generate more secure code without compromising functional correctness. Extensive evaluation across various scenarios shows that \toolname significantly improves the security of generated code while maintaining its functionality, and also demonstrates strong generalization across languages and models. This work takes a critical step toward securing LLM-based software development.  
% All data and code in this study are publicly available at:\textbf{\url{https://zenodo.org/records/15205184}}.

\begin{acks}

\end{acks}

\balance
\bibliographystyle{ACM-Reference-Format}%ACM-Reference-Format reference style
\bibliography{bib/references}

\newpage
\appendix
\section{Security Knowledge Example}
\begin{figure}[H]
    \centering
    \includegraphics[width=0.9\linewidth]{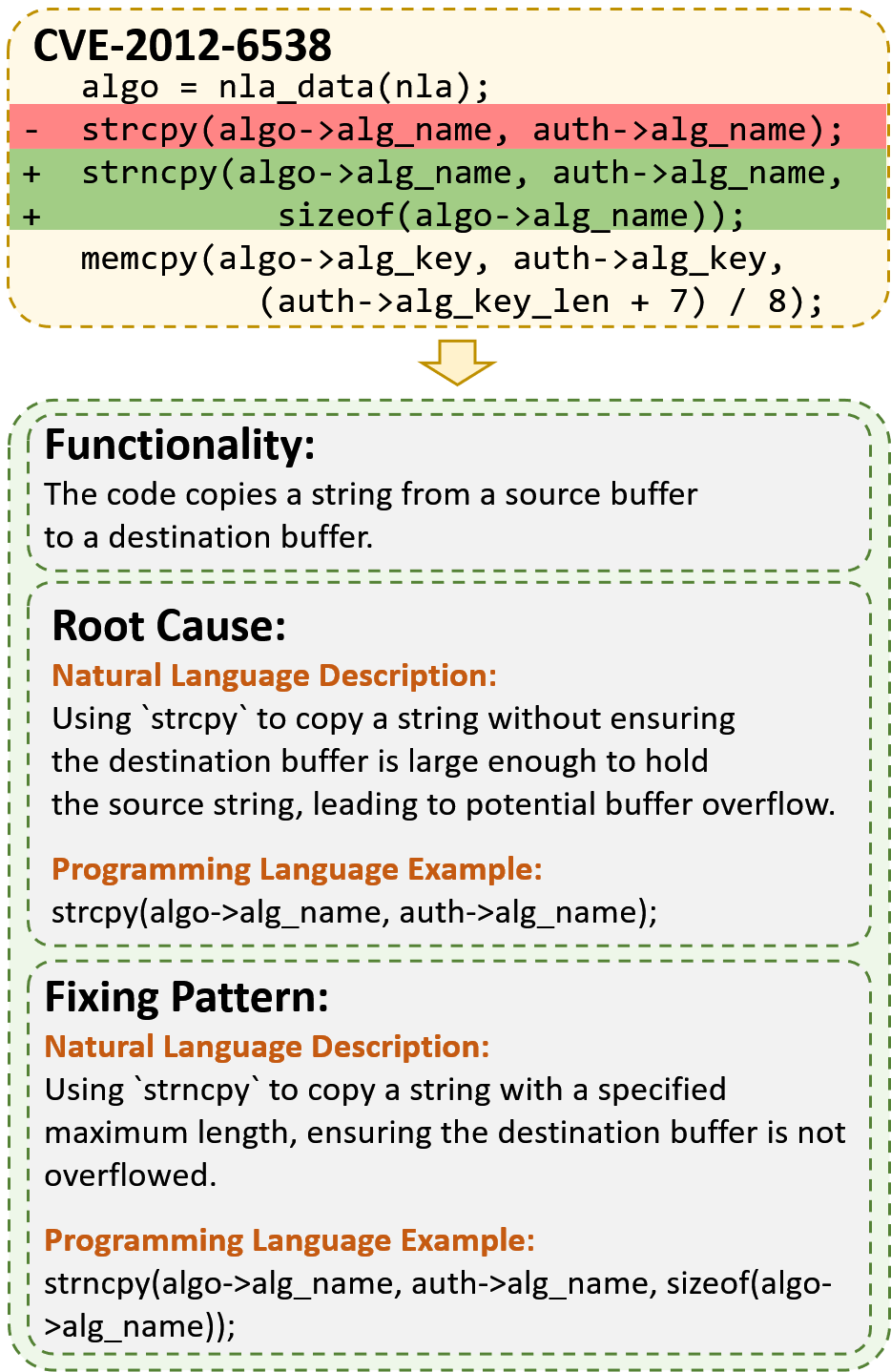}
    \caption{An example of security knowledge extracted from vulnerability}
    \label{fig:security_kng}
\end{figure}
Figure~\ref{fig:security_kng} presents an example of security knowledge extracted from CVE-2012-6538. This vulnerability arises from the use of an insecure function, which may lead to a buffer overflow. The patched code mitigates this issue by replacing {\tt strcpy} with {\tt strncpy}, enforcing a specific size constraint on the destination buffer.

This vulnerability arises from the use of an insecure function, which may lead to a buffer overflow. The patched code mitigates this issue by replacing {\tt strcpy} with {\tt strncpy}, enforcing a specific size constraint on the destination buffer. The \textbf{Functionality} dimension describes the fundamental operation of the vulnerable code snippet (\ie copying a string from a source buffer to a destination buffer). The \textbf{Root Cause} dimension provides a detailed explanation of the vulnerability in natural language along with an illustrative code example (\ie the risks associated with using {\tt strcpy}). Finally, the \textbf{Fixing Pattern} dimension includes both a description of the secure coding practice and an example of the corrected code (\ie replacing {\tt strcpy} with a safer alternative, {\tt strncpy}). This structured knowledge extraction process informs secure code generation in RACG, enabling \toolname to enhance the security of the generated code.

\section{Prompt Templates}
\subsection{Security Knowledge Extraction}
\begin{figure}[H]
    \centering
    \prompt{Prompt 1: Security Knowledge Extraction}{1}{ 
% I will provide a fixing commit for a \{$D_{cve}$\} vulnerability, along with its vulnerability description and commit details.
% Please help me describe the functionality of the vulnerable code snippet then identify and extract the root cause of the vulnerability introducing and corresponding fixing patterns.
% \textbf{\# Vulnerability Description}\\
% \{$D_{cve}$\}

% \textbf{\# Fixing Commit}\\
% \{DIFF\}

% \textbf{\# Output}\\
% Please output the following information in a JSON object format:
% \{"Functionality": a string that describes the functionality of the vulnerable code.
% "Root\_Cause": \{"Description": a string that describes the vulnerability's root cause. "Example": a code example illustrating the vulnerability.\}
% "Fixing\_Pattern": \{"Description": a string that describes the fixing pattern of the given vulnerability. "Example": a code example illustrating the fixed code of the given vulnerability.\}
% \}
\textbf{Task:} Analyze a vulnerability fixing commit to extract security knowledge.

    \textbf{Input:}
    \begin{itemize}
        \item \textbf{Vulnerability Description:} \{$D_{cve}$\}
        \item \textbf{Vulnerability Type:} \{$D_{cwe}$\}
        \item \textbf{Fixing Commit (Diff):} \{DIFF\}
    \end{itemize}
    \textbf{Instructions:}
    \begin{enumerate}
        \item Describe the functionality of the vulnerable code snippet.
        \item Identify and extract the root cause of the vulnerability.
        \item Identify and extract the corresponding fixing pattern.
    \end{enumerate}
    \textbf{Output Format:}
Provide the output in JSON format, adhering to the following structure:\\
\noindent \{ \\
        \makebox[1em][l]{} "Functionality": "<Description of the vulnerable code's functionality>", \\
        \makebox[1em][l]{} "Root\_Cause": [ \\
        \makebox[2em][l]{} "<Detailed description of the vulnerability's root cause>", \\
        \makebox[2em][l]{} "<Code example illustrating the vulnerability>" ]\\
        % \makebox[1em][l]{} ], \\
        \makebox[1em][l]{} "Fixing\_Pattern": [ \\
        \makebox[2em][l]{} "<Detailed description of the fixing pattern>", \\
        \makebox[2em][l]{} "<Code example illustrating the vulnerability repair>" ]\\
        % \makebox[1em][l]{} ] \\
        \noindent \}
}
\vspace{-2mm}
\end{figure}

In Prompt~\ref{prompt:1}, the terms $D_{\text{cve}}$ and $D_{\text{cwe}}$ denote the CVE description and the Common Weakness Enumeration (CWE) classification type, respectively. The \texttt{DIFF} represents the function-level diff, generated by comparing the vulnerable code $C_v$ with its corresponding fixed version $C_f$. Note that \texttt{DIFF} includes the full function context, as this detailed diff provides comprehensive information for understanding both the vulnerability and its resolution.

\subsection{Query Decomposition}
\begin{figure}[H]
    \centering
    \prompt{Prompt 2: Query Decomposition}{2}{ 
Please help me to break down a code generation query into smaller, detailed sub-tasks. For each sub-task, please utilize the natural language description of what the sub-tasks do, focusing on explaining the functionality of the sub-tasks.
\\
\textbf{\# User's Query}\\
\{QUERY\}

\textbf{\# Output}\\
Please output the following information in a JSON object format:
[\{"Description": "string"\}]

}
\vspace{-2mm}
\end{figure}

Prompt~\ref{prompt:2} shows the template used for query decomposition, where {\tt QUERY} represents the user's original query $Q$. 

\subsection{Security-Augmented Code Generation}
\begin{figure}[H]
    \centering
    \prompt{Prompt 3: Security-Augmented Code Generation}{3}{
        /*-----Original RACG Prompt Begin-----*/\\
        Write a C function that allocates memory for a string and copies its content, returning a pointer to the new string.\\
        
        \textbf{\# Code Examples}\\
        char* copy\_string(char* src) \{ \texttt{malloc}... \}\\
        /*-----Original RACG Prompt End-----*/\\
        \\
        /*-----Injected Security Knowledge-----*/ \\
        The code generation process involves the following sub-tasks. For each, I provide potential vulnerabilities and corresponding mitigation strategies.\\
        
        \textbf{\# Sub-Tasks with Corresponding Knowledge}\\
        \textbf{Sub-Task 1:} Copy each string from the input array to the corresponding location in the new array.\\
        \textbf{Security Knowledge 1:} ...\\
\noindent\{\\
\makebox[1em][l]{} "Functionality": ...,\\
\makebox[1em][l]{} "Root\_Cause": [ \\
        \makebox[2em][l]{} <Description>: Using \texttt{strcpy} to copy a string without ensuring the buffer is large enough. \\
        \makebox[2em][l]{} <Code Example>: \texttt{strcpy(...)};], \\
\makebox[1em][l]{} "Fixing\_Pattern": [ \\
        \makebox[2em][l]{} <Description>: Use \texttt{strncpy} to ensure the string is copied safely.\\
        \makebox[2em][l]{}<Code Example>: \texttt{strcnpy(...)};]\\
        % \makebox[1em][l]{} ] \\
\noindent\}
\\
...\\
        \textbf{Sub-Task N:}\{SUB\_TASK\_N\}\\
        \textbf{Security Knowledge N:}\{SEC\_KNOW\_N\}
    }
\end{figure}
Prompt~\ref{prompt:3} provides a concrete example of prompt for security-augmented code generation. The injected security knowledge includes the relevant knowledge retrieved for the given task. In the prompt, \{SUB\_TASK\_N\} and \{SEC\_KNOW\_N\} refer to the N-th sub-task and its corresponding retrieved security knowledge, respectively.

% that's all folks
\end{document}